\documentclass[traditabstract]{aa} % for the abstract without structuration % (traditional abstract) 
\usepackage[pdftex]{graphicx}
\usepackage{txfonts}
\usepackage{natbib}
\usepackage{longtable}
\def\ms{\hbox{\,m\,s$^{-1}$}}         %m.s -1
\def\m2s2{\hbox{\,m$^{2}$\,s$^{-2}$}} %m2.s -2
\def\kms{\hbox{\,km\,s$^{-1}$}}       %km.s -1
\def\Msun{\hbox{$\mathrm{M}_{\odot}$}}             %Msun
\def\Rsun{\hbox{$\mathrm{R}_{\odot}$}}
\def\Mjup{\hbox{$\mathrm{M}_{\rm Jup}$}}
\def\Rjup{\hbox{$\mathrm{R}_{\rm Jup}$}}

\def\mp{M_{\rm p}}
\def\rp{R_{\rm p}}

\begin{document}

\title{Improved parameters of seven Kepler giant companions characterized with SOPHIE and HARPS-N}
\titlerunning{Improved parameters of seven Kepler giant companions}
\authorrunning{Bonomo et al. 2014}

\author{A.~S.~Bonomo \inst{1} 
\and A.~Sozzetti\inst{1}
\and A.~Santerne\inst{2}
\and M.~Deleuil \inst{3}
\and J.-M.~Almenara \inst{3}
\and G.~Bruno \inst{3}  
\and R.~F.~D\'iaz\inst{4}
\and G.~H\'ebrard\inst{5, 6}
\and C.~Moutou\inst{3} 
}

\institute{
INAF - Osservatorio Astrofisico di Torino, via Osservatorio 20, 10025 Pino Torinese, Italy
\and Instituto de Astrof\'isica e Ci\^{e}ncias do Espa\c co, Universidade do Porto, CAUP, Rua das Estrelas, PT4150-762 Porto, Portugal 
\and Aix Marseille Universit\'e, CNRS, LAM (Laboratoire d'Astrophysique de Marseille) UMR 7326, 13388, Marseille, France
\and Observatoire Astronomique de l'Universit\'e de Gen\`eve, 51 chemin des Maillettes, 1290 Versoix, Switzerland
\and Observatoire de Haute-Provence, Universit\'e Aix-Marseille \& CNRS, F-04870 St.~Michel l'Observatoire, France
\and Institut d'Astrophysique de Paris, UMR7095 CNRS, Universit\'e Pierre \& Marie Curie, 98bis boulevard Arago, 75014 Paris, France
}

\offprints{A.~S.~Bonomo\\
 \email{bonomo@oato.inaf.it}}

\date{Received 13 November 2013 / Accepted 2 January 2015}

\abstract{Radial-velocity observations of \emph{Kepler} candidates obtained with the SOPHIE and HARPS-N spectrographs 
have permitted unveiling the nature of the five giant planets Kepler-41b, Kepler-43b, Kepler-44b, Kepler-74b, 
and Kepler-75b, the massive companion Kepler-39b, and the brown dwarf KOI-205b. 
These companions were previously characterized with long-cadence (LC) \emph{Kepler} data. 
Here we aim at refining the parameters of these transiting systems 
by \emph{i)} modelling the published radial velocities (RV) and \emph{Kepler} 
short-cadence (SC) data that provide a much better sampling of the transits,  
\emph{ii)} performing new spectral analyses of the SOPHIE and ESPaDOnS 
spectra, after improving our procedure for selecting and 
co-adding the SOPHIE spectra of faint stars ($K_{\rm p} \gtrsim 14$), and iii)
improving stellar rotation periods hence stellar age estimates through gyrochronology, 
when possible, by using all the available LC data up to quarter Q17.
Posterior distributions of the system parameters were derived 
with a differential evolution Markov chain Monte Carlo approach.
Our main results are as follows: 
\emph{a)} Kepler-41b is significantly larger and less dense than previously found because
a lower orbital inclination is favoured by SC data. This also affects the determination 
of the geometric albedo that is lower than previously derived: $A_{\rm g} < 0.135$;
\emph{b)} Kepler-44b is 
moderately smaller and denser than reported in the discovery paper, as a consequence of 
the slightly shorter transit duration found with SC data;
\emph{c)} good agreement was achieved with published Kepler-43, Kepler-75, and KOI-205 system parameters, 
although the host stars Kepler-75 and KOI-205 were found to be slightly richer in metals and hotter, respectively;
\emph{d)} the previously reported non-zero eccentricities of Kepler-39b and Kepler-74b might be spurious.
If their orbits were circular, the two companions would be smaller and 
denser than in the eccentric case. The radius of Kepler-39b is still larger than predicted 
by theoretical isochrones. Its parent star is hotter and richer in metals than previously determined.
}


% 5 {} token are mandatory

\maketitle

%
%________________________________________________________________

\section{Introduction}
\label{introduction}
Thanks to unprecedented photometric precision
and temporal coverage, the \emph{Kepler} space telescope has discovered
over two thousand small-sized planetary candidates with radii $R_{\rm p} < 4~R_{\oplus}$
\citep{Burkeetal2014}. 
At the same time, it has provided the exoplanet community with 
more than two hundred Jupiter-sized candidates, thus 
triggering further studies on the structure, formation, and evolution of giant companions 
as well as on their atmosphere, if the optical and/or 
nIR occultations are observed. 

Since 2010 we have been following up several \emph{Kepler} giant candidates 
orbiting faint stars with \emph{Kepler} magnitudes $K_{\rm p} \gtrsim 14$
using the SOPHIE spectrograph at the Observatoire de Haute Provence (France). 
In addition to determining the fraction of false positives among the \emph{Kepler}
giant candidates \citep{Santerneetal2012},
this intensive follow-up
allowed us to characterize 
the giant planets
Kepler-41b/KOI-196b \citep{Santerneetal2011a}, 
Kepler-43b/KOI-135b and Kepler-44b/KOI-204b \citep{Bonomoetal2012a}, 
Kepler-74b/KOI-200b and Kepler-75b/KOI-889b \citep{Hebrardetal2013};
the massive companion Kepler-39b/KOI-423b in the brown-dwarf desert \citep{Bouchyetal2011}, 
which could be either an extremely massive planet or a low-mass brown dwarf; 
and the brown dwarf KOI-205b \citep{Diazetal2013}. 
For two planets, that is for Kepler-74b and Kepler-75b, additional radial-velocity measurements
were taken with the HARPS-N spectrograph \citep{Cosentinoetal2012}, which has been installed at the 
Telescopio Nazionale Galileo at La Palma island in Spring 2012 (see \citealt{Hebrardetal2013}).

All these giant companions were characterized using 
Kepler data with long-cadence (LC) temporal sampling of 29.42~min,
usually because short-cadence (SC) photometric measurements, that is one point every 58~s, 
were not available at the moment of publication. 
However, the long-cadence sampling presents 
the strong inconvenience of distorting the transit shape. This effect leads to
longer transit durations, more V-shaped transits,
hence lower ratios between the semi-major axis and the stellar radius than the true ones. 
This yields lower stellar densities from the third Kepler law and thus makes 
both stellar and planetary radii 
appear larger than they actually are \citep{Kipping10}.
To overcome this problem, \citet{Kipping10} suggested to 
perform the transit fitting by oversampling the transit model and then 
binning the model samples to those of the LC before computing the 
chi-square or the likelihood function. His Eq.~(40) suggests a simple way to choose 
the resampling resolution, given the photometric precision of the light curve.
Following this prescription, \citet{KippingBakos11} analysed 
the LC data of Kepler-4b through Kepler-8b  using a bin number of 4.

When modelling the transits of the Kepler planets observed with 
SOPHIE and HARPS-N, we therefore followed this suggestion by 
\citet{Kipping10} and oversampled the transit model by a factor
five, which is higher than recommended by his Eq.~(40). 
However, in some cases, the analysis of short-cadence
data is mandatory especially
when the orbital period is close to an integer multiple of the LC sampling 
$\delta T_{\rm lc}$ because this prevents the transit from being well sampled in orbital phase.
The most evident case is Kepler-43b, whose orbital period  
is $P=147.99 \cdot  \delta T_{\rm lc}$ (see Fig.~5 in \citealt{Bonomoetal2012a}). 
In addition, the massive companion Kepler-39b did not present an optimal coverage of the transit
ingress and egress in the quarters Q1 and Q2 that were analysed by \citet{Bouchyetal2011}
because $P=1032.14 \cdot \delta T_{\rm lc}$ (see Fig.~9 in \citealt{Bouchyetal2011}). 

Moreover, by performing an homogeneous analysis of transit photometry from space and
oversampling the \emph{Kepler} LC data by a factor of ten instead of five, 
\citet{Southworth2012} derived smaller stellar and planetary radii for 
some of the aforementioned giant companions, although his results agree with ours within $2~\sigma$.
This would indicate that an oversampling of the transit model by a factor of five 
might not be ideal in all cases, hence an independent analysis
of SC data is certainly recommended. Nevertheless,  
some of the slightly different results obtained by \citet{Southworth2012} 
are also due to a better ephemeris and 
transit signal-to-noise ratio (S/N) because, in most cases, he used 
longer temporal series than we did, up to quarter Q6.

In this paper, we report the results of our analysis of 
the \emph{Kepler} SC data of Kepler-39, Kepler-41, Kepler-43, Kepler-44, 
Kepler-74, Kepler-75, and KOI-205, 
along with the previously published radial velocities (RV). 
\emph{Kepler} LC data up to quarter Q17 were used to refine stellar rotation periods 
by means of both generalised Lomb-Scargle periodograms \citep{ZechmeisterKurster2009}
and autocorrelation functions, 
when an unambiguous peak with $\rm FAP < 0.01 \%$ could be identified.
This allows us to estimate system ages through 
gyrochronology \citep{MamajekHillenbrand2008}, after 
deriving the B-V index colour and its uncertainty 
from Eq. (3) in \citet{SekiguchiFukugita00}.
The IDs and \emph{Kepler} magnitudes of the parent stars are listed in Table~\ref{startable_KOI}.
This work aims at refining the characterization of these systems and possibly
clarifying the apparently unusual properties of Kepler-39b and Kepler-41b. 
Indeed, the former was found to have a larger radius than predicted by theoretical
isochrones of \citet{Baraffeetal2003}, and \citet{Bouchyetal2011} were
unable to find any reasonable explanation for this behaviour.
The latter seemed to be a non-inflated planet despite its proximity to
the host star, and to occupy an atypical position in the radius-mass 
and radius-$T_{\rm eq}$ diagrams
of giant planets (see Figs.~9 and 10 in \citealt{Santerneetal2011a}). 

Moreover, we performed new 
spectral analyses of the planet-hosting stars after improving our 
procedure of selecting, treating, and co-adding the SOPHIE spectra.
A revision of the atmospheric parameters 
may have a significant impact on stellar, hence planetary, parameters.

We recognize the merit of an approach to revisiting stellar {\em or} planetary parameters of transiting systems 
that encompasses much larger samples  (e.g., \citealt{Torresetal2012, Southworth2012}) than the one presented here. Our work differs 
in that it performs a self-consistent re-analysis taking into account both 
photometric {\em and} spectroscopic measurements and constraints within a coherent Bayesian framework to derive the posterior density 
distributions of the full set of system parameters.

%%%%%%%%%%%%%%%%%%%
%%%%%%%%%%%%%%%%%%%%
%%%%%%%%%%%%%%%%%%%
\begin{table*}
\centering
\caption{IDs, coordinates, and magnitudes of the planet-hosting stars 
Kepler-39, Kepler-41, Kepler-43, Kepler-44, Kepler-74, Kepler-75, and KOI-205}
\renewcommand{\footnoterule}{}     
\begin{tabular}{lcccc}       
\hline\hline                 
\emph{Kepler} name & Kepler-39 & Kepler-41 & Kepler-43 & Kepler-44 \\
Object                       & KOI-423   & KOI-196   & KOI-135    & KOI-204  \\
\emph{Kepler} ID      & 9478990   & 9410930  & 9818381   &  9305831 \\
2MASS ID   & 19475046+4602034 & 19380317+4558539 & 19005780+4640057   &  20002456+4545437 \\
\emph{Kepler} magnitude $ \rm K_{p}$ & 14.33 & 14.46 & 13.96 & 14.68 \\
\hline
\\
\hline
\emph{Kepler} name & Kepler-74 & Kepler-75 & -  \\
Object & KOI-200 & KOI-889 & KOI-205  \\
\emph{Kepler} ID & 6046540  & 757450 & 7046804\\
2MASS ID   &  19322220+4121198  & 19243302+3634385 & 19415919+4232163\\
\emph{Kepler} magnitude $ \rm K_{p}$ & 14.41 & 15.26 & 14.52 \\
\hline\hline
\end{tabular}
\label{startable_KOI}      
\end{table*}

%~\\

\section{Data}
\subsection{Kepler photometry}
\label{kepler_photometry}
Short-cadence measurements obtained with 
the simple-aperture-photometry 
pipeline\footnote{http://keplergo.arc.nasa.gov/PyKEprimerLCs.shtmlp} 
\citep{Jenkinsetal2010} were downloaded 
from the MAST archive\footnote{http://archive.stsci.edu/kepler/data\_search/search.php}. 
Eleven quarters of SC data (Q3-Q7 and Q10-Q15)  are available for Kepler-43; 
six quarters (Q10-Q15) for Kepler-75; four quarters (Q4-Q7) for Kepler-41, Kepler-44, Kepler-74, and KOI-205;
and three quarters (Q12-Q14) for Kepler-39.

The medians of the errors of SC measurements of Kepler-39, Kepler-41, Kepler-43, 
Kepler-44, Kepler-74, Kepler-75 and KOI-205 are $1.20 \cdot 10^{-3}$, $1.27 \cdot 10^{-3}$, 
$8.9 \cdot 10^{-4}$, $2.02 \cdot 10^{-3}$, $1.24 \cdot 10^{-3}$, $3.26 \cdot 10^{-3}$, and 
$1.28 \cdot 10^{-3}$ in units of relative flux, respectively. 

For all targets, the flux excess originating from background stars that are located within 
the \emph{Kepler} photometric mask was subtracted separately for each quarter by using 
the estimates provided by the \emph{Kepler} team\footnote{http://archive.stsci.edu/kepler/kepler\_fov/search.php; 
for Season 2 data of KOI-205, the value of crowding factor as derived by \citet{Diazetal2013} was used
(see \citealt{Diazetal2013}).}.
Indeed, this contamination of the target flux dilutes the transits, making them appear shallower
than they are, even though it usually does not exceed 5-7\% of the total collected flux.

All the SC data were used to model the transits of the seven giant companions.
The signal-to-noise ratios of the phase-folded transits are of at least $\sim 450$ and, in some cases, 
higher than 1\,000. 
Thanks to these high S/N, we were able to derive stellar and planetary radii with uncertainties 
$\lesssim 3\%$ almost in all cases (see Tables~\ref{table_param_KOI423}-\ref{table_param_KOI205}).
At this level of precision, errors on planetary radii are dominated by the
uncertainties on stellar models \citep{Southworth2011, Southworth2012} 
and/or on the orbital eccentricity $e$ and argument of periastron $\omega$. 
Indeed, the uncertainties on $e$ and $\omega$ from RV observations propagate into 
the transit parameter $a/R_\star$ and thus into the stellar density from the third Kepler law. 
Stellar density is then used as a proxy for luminosity 
to determine stellar hence planetary parameters (e.g., \citealt{Sozzettietal2007}) 
when no constraints from asteroseismology are available. 
This means that in our particular cases the additional use of LC data practically 
does not yield any significant improvement on planetary parameters 
while introducing possible covariances between transit parameters \citep{PriceRogers2014}.
For this reason, as mentioned before, we used LC data only to derive stellar rotation periods.

\subsection{Radial-velocity data}
\label{RV_data}
The RV observations considered in this work are those listed
in the announcement papers 
because no additional observations with either SOPHIE 
or HARPS-N were carried out for these targets.
The SOPHIE measurements were performed in
high-efficiency mode with a resolution of $\sim 40\,000$, and
exposure times not exceeding 1~hr.
The observations of KOI-200 and 
KOI-889 carried out with HARPS-N were taken in high-resolution mode 
(the only available one) with a resolving power
of $\sim 110\,000$ and exposures of 45~min and shorter than 25~min,
respectively.
Both SOPHIE and HARPS-N measurements were performed in  obj\_AB 
observing mode with fibre A centred on the target and fibre B on the sky.
When needed, the observations were 
corrected for moonlight pollution, as described in \citet{Bonomoetal2010}.

\subsection{Spectra}
\label{spectra}
The atmospheric parameters of the host stars, 
along with the stellar density derived from the transit fitting,
are of fundamental importance for determining 
stellar, hence planetary, parameters \citep{Sozzettietal2007}.

While radial-velocity measurements can accommodate low S/N spectra, 
spectral analysis is more challenging. 
Indeed, some diffuse light in the SOPHIE spectrograph 
might affect the spectra at very low S/N acquired in high-efficiency mode. 
For this reason, we recently improved our procedure of treating and selecting 
the SOPHIE spectra to determine stellar atmospheric parameters.
In particular, spectra with an S/N lower than 14 were excluded from the co-addition. 
Those acquired in the presence of the Moon were corrected for 
the moonlight contamination 
by subtracting the background as estimated from fibre B. 
As is usually done, the individual exposures were then set in the rest frame 
and co-added in a single master spectrum. 

In the case of Kepler-39, the co-added spectrum obtained 
this way shows deeper lines than simply co-adding 
all the SOPHIE spectra, as was previously done by \citet{Bouchyetal2011}. 
This has a significant impact on the 
derivation of the atmospheric parameters 
(see Sect.~\ref{results_atmos_param}).

The S/N of the SOPHIE master spectra at 600~nm and 
per element of resolution ranges between 100 and 170
for Kepler-39, Kepler-41, Kepler-43, Kepler-44, and Kepler-74. It is equal to 72 and 65 
for KOI-205 and Kepler-75, respectively.

Two host stars, namely KOI-205 and Kepler-39, were also observed 
with ESPaDOnS at the 3.6-m Canada-France-Hawaii Telescope in Mauna Kea 
as part of a programme dedicated 
to the characterization of Kepler 
planet-hosting stars\footnote{programme 12BF24, PI: M.~Deleuil}. 
The objective is indeed to carry out a better spectral analysis 
of the parent  stars with a spectrograph that offers both a
higher spectral resolution (R $\simeq$ 65 000) and
an extended spectral coverage (370 - 1000 nm).
These two targets were observed in 'object+sky' mode. 
The ESPaDOnS spectrum of KOI-205 with a $S/N \sim 90$ was previously used by 
\citet{Diazetal2013} to determine the host star and brown dwarf parameters. 
Kepler-39 was observed with ESPaDOnS on September 28, 2012 and 
December 1, 2012, in a series of five exposures of $\sim 40$~min.
The individual spectra as reduced by the CFHT Upena/Libre-Esprit pipeline were 
co-added after they were set in the rest frame and resulted  
in a master spectrum with S/N of 65 in the continuum at 600~nm
per resolution element. This co-added spectrum is analysed 
here for the first time.

\section{Data analysis}
\subsection{Spectral analysis}
\label{spectral_analysis}
To determine the effective temperature ($T_{\rm{eff}}$), surface gravity
($\log g$), and iron abundance [Fe/H], the co-added SOPHIE spectra
obtained with our improved selection and those acquired with
ESPaDOnS were reanalysed following the same procedures described in detail
by \citet{Sozzettietal2004, Sozzettietal2006} and references therein. A set of $\sim60$ relatively weak
lines of Fe I and 10 of Fe II were selected, and EWs
were measured using the TAME software \citep{KangLee2012}. Metal abundances
were derived assuming local thermodynamic equilibrium
(LTE), using the 2010 version of the spectral synthesis code MOOG 
\citep{Sneden1973}, a grid of Kurucz ATLAS plane-parallel model stellar atmospheres
\citep{Kurucz93}, and imposing excitation and ionisation equilibrium.
Uncertainties in the parameters were estimated following the
prescriptions of \citet{NeuforgeMagain1997} and \citet{GonzalezVanture1998}
and rounded to 25 K in $T_{\rm{eff}}$ and 0.05 dex in $\log g$.

The derived atmospheric parameters are compared in Sect.~\ref{results_atmos_param}
with the literature values, which were obtained with \emph{i)} the 
iterative spectral synthesis package VWA \citep{Brunttetal2010} for
Kepler-39, Kepler-41, Kepler-74, Kepler-75, and KOI-205, 
and \emph{ii)} the 2002 version of the MOOG code 
with the methodology described in \citet{Bonomoetal2012a} 
and \citet{Mortieretal2013} for Kepler-43 and Kepler-44.

\subsection{Combined analysis of Kepler and radial-velocity data}
\label{modelling}
To derive system parameters, 
a Bayesian analysis of \emph{Kepler} SC photometry and 
radial-velocity measurements was performed, 
using a differential evolution Markov chain Monte Carlo (DE-MCMC) method
\citep{TerBraak2006, Eastmanetal2013}. 
For this purpose, the epochs of the SOPHIE and HARPS-N observations 
were converted from $\rm BJD_{UTC}$ 
into $\rm BJD_{TDB}$ \citep{Eastmanetal2010}, which is the 
time stamp of \emph{Kepler} data. 

The transit fitting was performed using the model 
of \citet{Gimenez06, Gimenez09}. For this purpose, 
each transit was normalised by locally 
fitting a slope to the light-curve intervals of twice the transit 
duration before its ingress and after its egress. For Kepler-39, a linear function
of time did not provide a satisfactory normalisation because of the short-term 
stellar variability (see Sect.~\ref{system_param}), hence a quadratic function 
of time was used. Correlated noise was estimated following 
\citet{Pontetal2006} and \citet{Bonomoetal2012b}, and added in quadrature to 
the formal error bars. However, it turned out to be very low, generally 
lower than one fifth of the formal photometric errors, as expected for 
high-precision space-based photometry 
(e.g., \citealt{Aigrainetal2009}, \citealt{Bonomoetal2012b}). 

Our global model has 12 free parameters when i) an eccentric model
was considered and ii) RV were taken with only one instrument (SOPHIE): 
the transit epoch $T_{\rm 0}$; the orbital period $P$; 
the systemic radial velocity $V_{\rm r}$; the radial-velocity semi-amplitude $K$; 
$\sqrt{e}~{\cos{\omega}}$ and $\sqrt{e}~{\sin{\omega}}$
(e.g., \citealt{Andersonetal2011});
an additive RV jitter term $s_{\rm j}$ to account for possible jitter in the RV measurements 
regardless of its origin, such as instrumental effects, 
stellar activity, additional companions, etc.; 
the transit duration from first to fourth contact $T_{\rm 14}$;
the ratio of the planetary-to-stellar radii $R_{\rm p}/R_{*}$;
the inclination $i$ between the orbital plane and the plane of the sky;
and the two limb-darkening coefficients (LDC)
$q_{1}=(u_{a}+u_{b})^2$ and $q_{2}=0.5 u_{a} / (u_{a}+u_{b})$ \citep{Kipping2013}, 
where $u_{\rm a}$ and $u_{\rm b}$ 
are the coefficients of the limb-darkening quadratic law\footnote{
$I(\mu)/I(1)=1-u_{\rm a}(1-\mu)-u_{\rm b}(1-\mu)^2$, where $I(1)$ is the 
specific intensity at the centre of the disc and $\mu=\cos{\gamma}$, 
$\gamma$ being the angle between the surface normal and the line of sight. }.
Two additional parameters, that is the HARPS-N systemic radial velocity and jitter term, 
were fitted when HARPS-N data were obtained as well (Kepler-74 and Kepler-75). 
Uniform priors were set on all parameters, in particular with bounds of [0, 1] for  
$q_{1}$ and $q_{2}$ \citep{Kipping2013}, lower limit of zero for 
$K$ and $s_{\rm j}$, and upper bound of 1 for $e$ (the lower limit 
of 0 simply comes from the choice of fitting $\sqrt{e}~{\cos{\omega}}$ and 
$\sqrt{e}~{\sin{\omega}}$). 

The posterior distributions of our free parameters 
were determined by means of our DE-MCMC code by 
maximising a Gaussian likelihood (see, e.g., Eq.~9 and 10
in \citealt{Gregory2005}).
For each target, a number of chains equal to twice the number of 
free parameters were run simultaneously
after being started at different positions in the parameter space 
but reasonably close to the system values 
known in the literature and/or obtained with an independent 
fit that was previously performed with AMOEBA \citep{NelderMead1965}. 
The jumps for a current chain in the parameter space  
were determined from the other chains, according 
to the prescriptions given by \citet{TerBraak2006}, 
and the Metropolis-Hastings algorithm 
was used to accept or reject a proposed step for each chain. 

For the convergence of the chains, we required  
the Gelman-Rubin statistics, \emph{\^{R}}, to be lower than 1.03 for all the 
parameters \citep{Gelmanetal2003}. Steps belonging to the burn-in phase were identified 
following \citet{Knutsonetal2009} and were excluded.
The medians of the posterior distributions of the 
fitted and derived parameters and their $34.13\%$ intervals
are reported as the final values and their $1\sigma$ error bars. 
When the distributions of the eccentricity and the RV jitter were found to
peak at zero, we provided only the $1\sigma$ upper limits estimated as 
the $68.27\%$ confidence intervals starting from zero. 
Indeed, the medians of these distributions
might yield misleading non-zero values.

Finally, the Yonsei-Yale evolutionary tracks \citep{Demarqueetal2004} 
for the effective temperature, metallicity, and density 
of the host stars were used to determine the stellar, hence companion,
parameters \citep{Sozzettietal2007, Torresetal2012}.

For Kepler-41 and Kepler-74 only a circular model was adopted 
for the following reasons: 
the RV curve of Kepler-41 is slightly asymmetric 
with respect to a sinusoid, very likely because of residual effects 
from the correction of moonlight contamination and/or
low S/N (from 13 to 20) spectra. These artificial asymmetries in the 
RV curve tend to bias the solution of system parameters towards 
a low but non-zero eccentricity $e \sim 0.14$ even when including 
in the global fit the secondary eclipse 
that indicates that $e~\cos{\omega}$ is consistent 
with zero (e.g., \citealt{Santerneetal2011a, Quintanaetal2013}). 
This would reduce the orbital configurations of a possible eccentric 
orbit to $\omega=90$ or 270~deg.
However, the expected circularization timescale is shorter than
$100$~Myr by assuming a modified tidal quality factor of 
$Q^{'}_{\rm p}=10^7$ for the planet 
because of its short orbital period $P=1.85$~days (hence small semi-major axis 
$a=0.031$~au) and relatively low mass $\mp \sim 0.6~\Mjup$ for a 
Jupiter-sized planet. Indeed, there are no planets with mass comparable 
to Kepler-41 and $P < 3$~d with a significant eccentricity. 
For these reasons, we adopted a circular model
for Kepler-41.

For Kepler-74, our DE-MCMC chains did not converge 
towards a unique solution when we varied the eccentricity, which resulted in 
very low acceptance rates. 
This was also noticed by \citet{Hebrardetal2013}, who imposed a 
Gaussian prior on the orbital eccentricity, solely based on a RV fit, 
in their combined analysis of \emph{Kepler} and RV data.
However, this prior inevitably affects the posterior distributions of orbit
and transit parameters. Instead, we preferred to use a circular model given that
current RV data evidently do not allow us to constrain the orbital eccentricity well.

The Kepler-39 system parameters were obtained by using both 
eccentric and circular models because the $2\sigma$ significance 
of the eccentricity $e=0.112 \pm 0.057$ cannot exclude that 
the orbit is perfectly circular \citep{LucySweeney1971}.

\section{Results}
\label{results_analysis}

\subsection{Stellar atmospheric parameters}
\label{results_atmos_param}
The $T_{\rm eff}$ and [Fe/H] of Kepler-41, Kepler-43, Kepler-44, and Kepler-74, which 
were determined with the procedure described in Sect.~\ref{spectral_analysis}, 
are consistent within $1~\sigma$ with the literature values. 
The surface gravities of Kepler-43 and Kepler-44
were found to be $\log g=4.4 \pm 0.10$
(Kepler-43) and $\log g=4.1 \pm 0.10$ (Kepler-44), which are lower than previously found
by \citet{Bonomoetal2012a}, that is 
$\log g=4.64 \pm 0.103$ and $4.59 \pm 0.14$, respectively.
These newly determined values are more consistent with 
the photometrically derived $\log g$ 
(see Tables~\ref{table_param_KOI135} and \ref{table_param_KOI204}).

The star KOI-205 was found to be slightly hotter than reported by \citet{Diazetal2013}, 
with $T_{\rm eff}=5400 \pm 75$~K, metallicity 
[Fe/H]=$0.18 \pm 0.12$, and $\log g=4.7 \pm 0.10$, 
both with the ESPaDOnS and the co-added SOPHIE spectrum 
(see Table~\ref{table_param_KOI205}). The slightly hotter temperature
has only a minor, almost negligible, influence on system parameters.

The most striking differences with previously determined atmospheric parameters 
were found for Kepler-39 and Kepler-75. The former is significantly richer 
in metals and slightly hotter than reported by \citet{Bouchyetal2011}: $T_{\rm eff}=6350 \pm 100$~K,
 $\rm [Fe/H]=0.10 \pm 0.14$, and $\log g=4.4 \pm 0.15$, to be compared with
the previous estimates $T_{\rm eff}=6260 \pm 140$, 
 $\rm [Fe/H]=-0.29 \pm 0.10$, and $\log g=4.1 \pm 0.2$. 
Very consistent values were derived from the analysis of the ESPaDOnS spectrum.
This difference comes from the better selection and treatment of the SOPHIE spectra 
before co-adding the individual spectra (see Sect.~\ref{spectra}). 
It is not due to the different spectral analysis technique 
that was used by \citet{Bouchyetal2011}, that is, the VWA package
\citep{Brunttetal2010}. Indeed, when run on the new co-added 
SOPHIE spectrum, VWA provided results that are almost identical to those obtained with MOOG:
$T_{\rm eff}=6360 \pm 100$~K, $\rm [Fe/H]=0.12 \pm 0.14$, and $\log g=4.4 \pm 0.3$.
These new values have an impact on stellar mass and age
(see Table~\ref{table_param_KOI423}).

Kepler-75 is also found to be 
richer in metals than previously thought. Its atmospheric parameters 
determined with MOOG are $T_{\rm eff}=5200 \pm 100$~K, 
$\rm [Fe/H]=0.30 \pm 0.12$, and $\log g=4.6 \pm 0.15$. 
While the $T_{\rm eff}$ is consistent within $1~\sigma$ with 
the value derived by \citet{Hebrardetal2013}, the difference in metallicity is 
about $2~\sigma$ (see Table~\ref{table_param_KOI889}). 
However, at these temperatures, the system parameters of Kepler-75 
do not significantly change as a consequence of the higher metallicity.

\subsection{System parameters}
\label{system_param}

\subsubsection{Kepler-39}
Orbital and transit parameters obtained with our Bayesian DE-MCMC analysis
in the eccentric case agree well with those that were previously determined by
\citet{Bouchyetal2011}. However, we found a lower significance of approximately
$2~\sigma$ for the orbital eccentricity, that is $e=0.112 \pm 0.057$, while \citet{Bouchyetal2011}
reported $e=0.122 \pm 0.023$. Our larger uncertainty on the eccentricity
indicates that it might be spurious, according to \citet{LucySweeney1971}. 
The Bayes factor of $\sim 3$ between the eccentric and the circular model, which was
computed by using the truncated posterior mixture method \citep{TuomiJones2012}, 
does not provide strong enough evidence either for an eccentric orbit, according to \citet{KassRaftery1995}.

As discussed in Sect.~\ref{results_atmos_param}, 
the stellar atmospheric parameters were refined thanks 
to both a better treatment of the SOPHIE spectra and a new 
ESPaDOnS spectrum. Specifically, a moderately hotter temperature 
$T_{\rm eff}=6350 \pm 100$ K and a significantly 
higher metallicity of $0.10 \pm 0.14$ were found (see Sect.~\ref{results_atmos_param}).
For these new atmospheric parameters and 
the transit density, the Yonsei-Yale evolutionary tracks
indicate a more massive and younger star: 
$M_\star=1.29_{-0.07}^{+0.06}~\Msun$, $R_\star=1.40 \pm 0.10~\Rsun$, and
age of $2.1_{-0.9}^{+0.8}$~Gyr.
The corresponding mass, radius, and density of Kepler-39b 
are $M_{\rm b}=20.1_{-1.2}^{+1.3}~\Mjup$, $R_{\rm b}=1.24_{-0.10}^{+0.09}~\Rjup$, and
$\rho_{\rm b}=13.0_{-2.2}^{+3.0}~\rm g\;cm^{-3}$.
These companion parameters are consistent with 
those reported by \citet{Bouchyetal2011}, except for the stellar age, 
which is about half the value found by these authors (see Table~\ref{table_param_KOI423}).
Interestingly, this updated value of the age agrees well 
with the gyrochronology estimate \citep{MamajekHillenbrand2008}, 
that is $t_{\rm gyr}=0.7_{-0.3}^{+0.9}$~Gyr, for the stellar rotation period $P_{\rm rot}=4.50 \pm 0.07$~d
inferred from the \emph{Kepler} LC light curve.

\begin{figure}[h!]
\centering
\includegraphics[width=7.5cm]{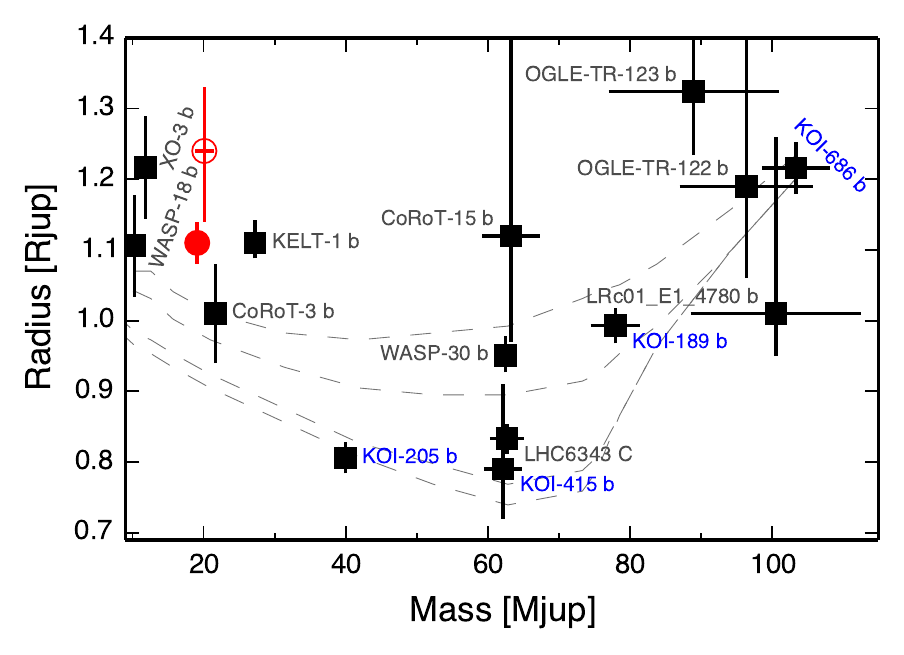}
\caption{Radius-mass diagram including transiting companions more massive
than $10~\Mjup$. The dashed curves are the \citet{Baraffeetal2003}
isochrones for 0.5, 1, 5, and 10 Gyr (from top to bottom).
Both circular (filled red circle) and eccentric (empty red circle) 
solutions are shown for Kepler-39b. Blue labels indicate the brown dwarfs that
were characterized thanks to SOPHIE spectroscopic measurements 
(see also \citealt{Moutouetal2013, Diazetal2014} ). }
\label{fig_MR_BD}
\end{figure}

Figure~\ref{fig_MR_BD} shows the position of Kepler-39b 
(empty red circle) in the radius-mass diagram of transiting companions
with masses between 10 and 100~$\Mjup$ for the eccentric case. The dashed lines
from top to bottom show the \citet{Baraffeetal2003} isochrones 
for 0.5, 1, 5, and 10~Gyr. As already noted by \citet{Bouchyetal2011}, the companion radius
given by the eccentric solution would be incompatible with 
theoretical isochrones with a probability $> 95\%$ ($2~\sigma$). 
However, the solution these authors proposed to explain the large radius of Kepler-39b, 
that is an increased opacity in the companion atmosphere,
which previously seemed unlikely for the low stellar metallicity, 
now might apply for Kepler-39b. Indeed, our new estimate of [Fe/H]  
is significantly higher. 

\begin{figure*}[t]
\centering
\begin{minipage}{15cm}
%\vspace{-0.5cm}
%\includegraphics[width=7cm]{plot_transit_KOI423_SC_circ_ecc_2014.pdf}
\includegraphics[width=7cm]{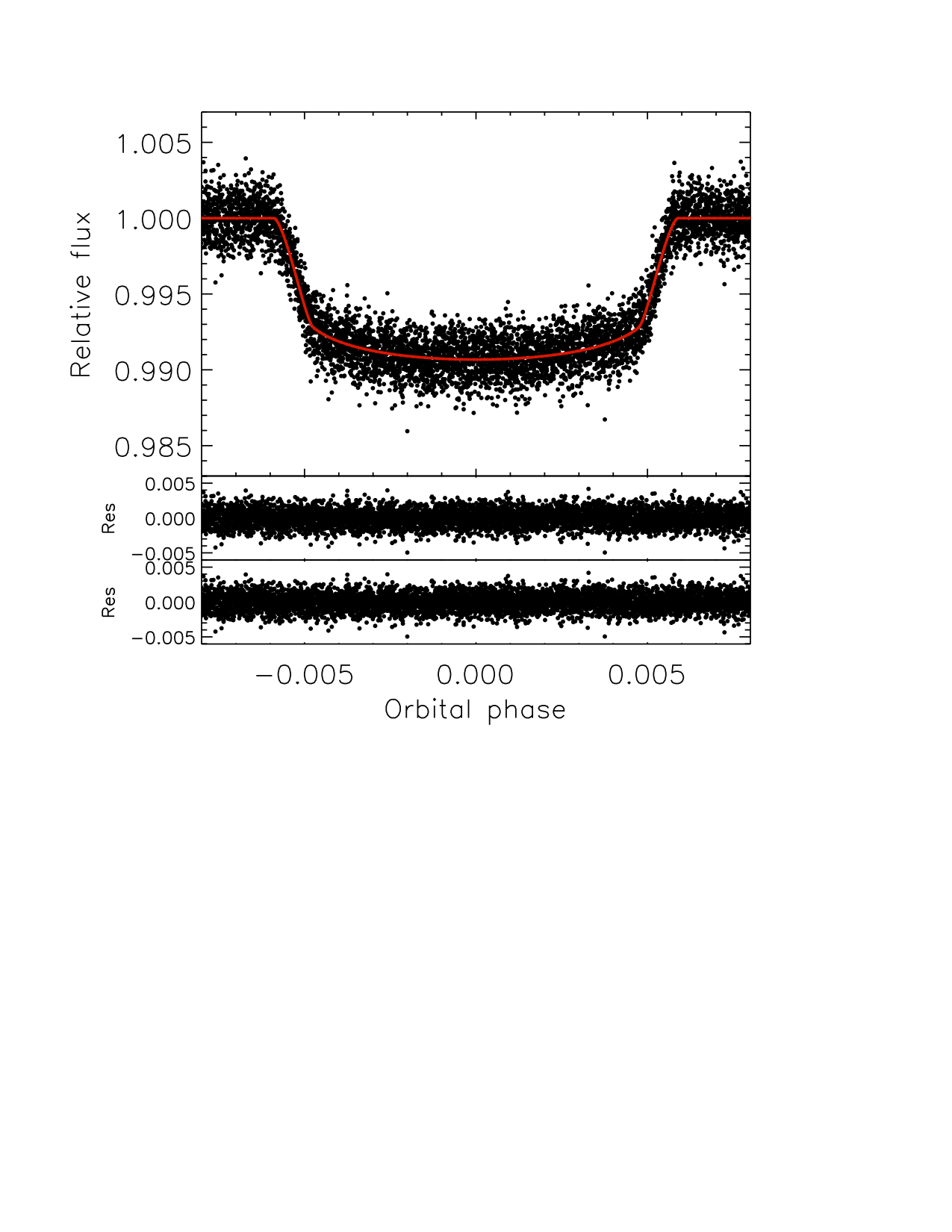}
\hspace{1.0cm}
\includegraphics[width=7cm]{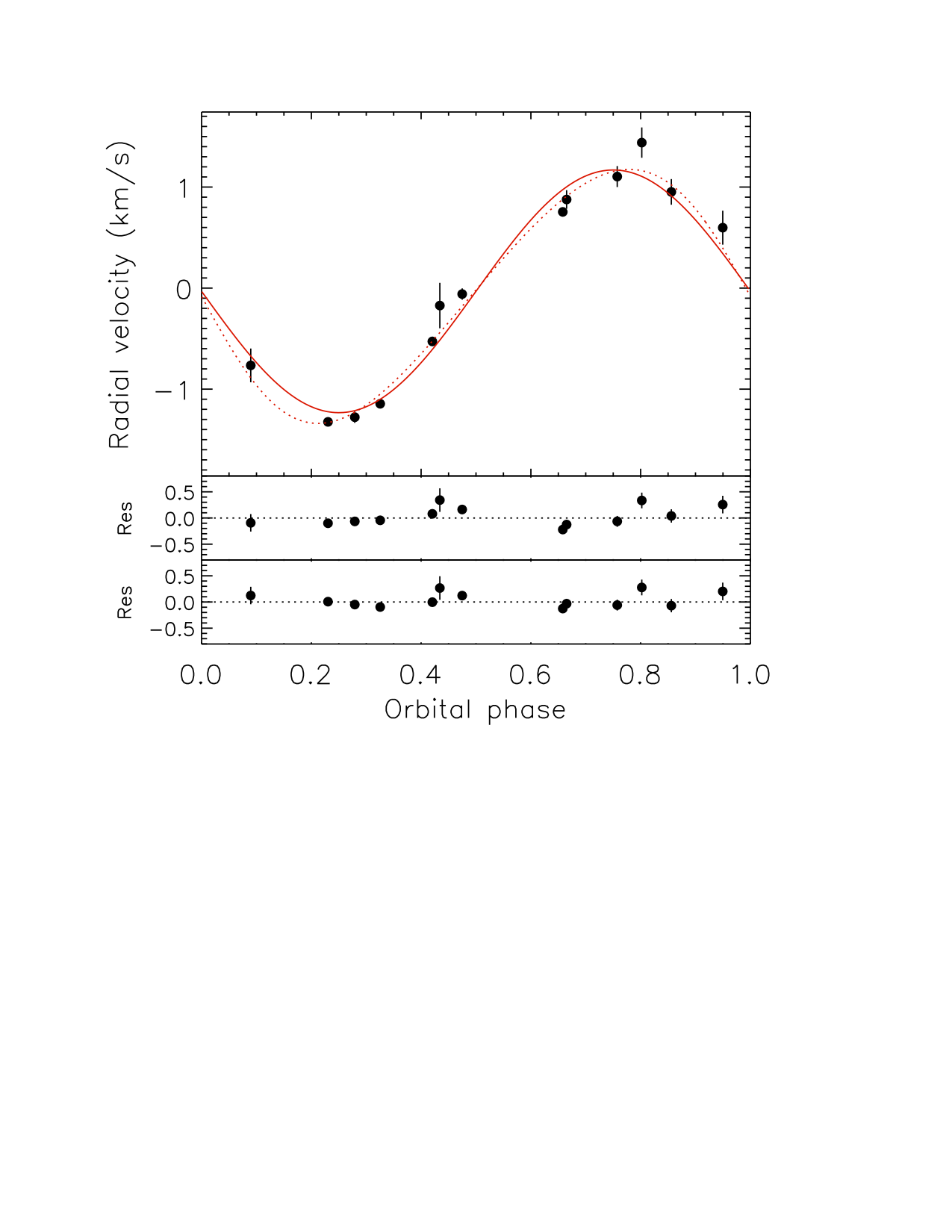}
\vspace{-3.7cm}
\caption{
\emph{Left panel.} \emph{Top}: phase-folded transit of Kepler-39b along with the transit
models for the circular (red solid line) and eccentric (red dotted line) orbits. The two models 
are indistinguishable. \emph{Middle}: residuals of the circular orbit. 
\emph{Bottom}: residuals of the eccentric orbit.
\emph{Right panel.} \emph{Top}: phase-folded radial-velocity curve of Kepler-39 and, superimposed, the 
Keplerian models for the circular (red solid line) and eccentric (red dotted line) orbits. 
\emph{Middle}: O-C of the circular orbit. \emph{Bottom}: O-C of the eccentric orbit.
}
\label{fig_KOI423}
\end{minipage}
\end{figure*}

As previously discussed, the eccentricity of Kepler-39b might be spurious  
according to the Lucy-Sweeney criterion. 
In the circular case, stellar and companion radii are slightly smaller than in the eccentric case, 
which implies a higher bulk density of $17.4_{-1.4}^{+1.6}$~$\rm g\;cm^{-3}$ 
(see Table~\ref{table_param_KOI423}).  
The system age would be $1.0_{-0.7}^{+0.9}$~Gyr. 
The position of Kepler-39b in the circular case is shown in Fig.~\ref{fig_MR_BD} 
with a red filled circle. Figure \ref{fig_KOI423} displays the phase-folded transit and
RV curve and the best solutions obtained with both eccentric and circular models.

\begin{table*}[h!]
\centering
\caption{Kepler-39 system parameters.}            
\begin{minipage}[t]{15.0cm} 
\setlength{\tabcolsep}{1.2mm}
\renewcommand{\footnoterule}{}                          
\begin{tabular}{l l l l}        
\hline\hline                 
\emph{Fitted system parameters}  & \citet{Bouchyetal2011} & This work (eccentric) & This work (circular) \\
\hline
Orbital period $P$ [days] & 21.0874 $\pm$ 0.0002 &  $21.087210 \pm  0.000037 $ &  $21.087212 \pm 0.000030 $  \\
Transit epoch $T_{ \rm 0} [\rm BJD_{TDB}-2454900$] & 72.5959 $\pm$ 0.0006 &  $1042.60708 \pm 0.00024 $  &  $1042.60707 \pm 0.00021 $ \\
Transit duration $T_{\rm 14}$ [h] & 6.02 $\pm$ 0.09 &  $5.960 \pm 0.020 $ &  $5.960 \pm 0.016 $ \\
Radius ratio $R_{\rm b}/R_{*}$ & $0.0896_{-0.0012}^{+0.0011}$ &  $0.0910_{-0.0008}^{+0.0006} $ &  $0.0911 \pm 0.0006 $  \\
Inclination $i$ [deg] & $88.83_{-0.40}^{+0.59}$ &  $89.07 \pm 0.22 $ &  $89.23_{-0.11}^{+0.13} $ \\
Limb-darkening coefficient $q_{1}$  &  - &  $0.23_{-0.06}^{+0.08} $ &  $0.22_{-0.05}^{+0.06} $ \\
Limb-darkening coefficient $q_{2}$  &  - &  $0.32_{-0.10}^{+0.14} $ &  $0.33_{-0.09}^{+0.12} $  \\
$\sqrt{e}~\cos{\omega}$ & - &  $-0.047_{-0.078}^{+0.084} $ & 0 (fixed) \\
$\sqrt{e}~\sin{\omega}$ & - &  $0.324_{-0.132}^{+0.080} $  & 0 (fixed) \\
Orbital eccentricity $e$  &  $0.122 \pm 0.023$ &  $0.112 \pm 0.057 $ & 0 (fixed)  \\
Argument of periastron  [deg] $\omega$  &  $98.9_{-6.8}^{+5.9}$ &  $99_{-14}^{+22} $ & 90 (fixed)  \\
Radial velocity semi-amplitude $K$ [\kms] & $1.251 \pm 0.030$ &  $1.257 \pm 0.064 $ &  $1.201 \pm 0.050 $ \\
Systemic velocity  $V_{\rm r}$ [\kms] & $-0.101_{-0.015}^{+0.017}$ &  $-0.063 \pm 0.044 $ &  $-0.032_{-0.037}^{+0.040} $ \\
RV jitter [\ms] $s_{\rm j}$ & - &  $108_{-40}^{+56} $ &  $140_{-33}^{+40} $ \\
& \\
\multicolumn{2}{l}{\emph{Derived transit parameters}} \\
\hline
$a/R_{*}$ & $23.8_{-1.7}^{+1.8}$ &  $24.92_{-1.5}^{+1.9} $ &  $27.74_{-0.49}^{+0.55} $ \\
Stellar density $\rho_{*}$ [$ \rm g\;cm^{-3}$] & $0.57_{-0.11}^{+0.14}$&  $0.66_{-0.11}^{+0.15} $ &  $0.91_{-0.05}^{+0.06} $ \\
Impact parameter $b$ & $0.43_{-0.18}^{+0.11}$ &  $0.36_{-0.08}^{+0.05} $ &  $0.37_{-0.06}^{+0.04} $ \\
Limb-darkening coefficient $u_{a}$  & $0.303 \pm 0.014$~$^a$ &  $0.31 \pm 0.07 $ &  $0.32 \pm 0.06 $ \\
Limb-darkening coefficient $u_{b}$  & $0.308 \pm 0.005$~$^a$  &  $0.17 \pm 0.14 $ &  $0.15 \pm 0.11 $ \\
& \\
\multicolumn{2}{l}{\emph{Atmospheric parameters of the star}} \\
\hline
Effective temperature $T_{\rm{eff}}$[K] & 6260 $\pm$ 140 & 6350 $\pm$ 100 & 6350 $\pm$ 100 \\
Spectroscopic surface gravity log\,$g$ [cgs]&  4.1  $\pm$ 0.2 & 4.40  $\pm$ 0.15 & 4.40  $\pm$ 0.15 \\
Derived surface gravity log\,$g$ [cgs]&  4.19  $\pm$ 0.07 &  $4.25 \pm 0.06 $ &  $4.34  \pm 0.02 $ \\
Metallicity $[\rm{Fe/H}]$ [dex] & -0.29  $\pm$ 0.10 & 0.10  $\pm$ 0.14 & 0.10  $\pm$ 0.14 \\
Stellar rotational velocity $V \sin{i_{*}}$ [\kms] & 16 $\pm$ 2.5 & 16 $\pm$ 2.5 & 16 $\pm$ 2.5\\
Spectral type & F8IV & F7V & F7V \\
& \\
\multicolumn{2}{l}{\emph{Stellar and planetary physical parameters}} \\
\hline
Stellar mass [\Msun] &  $1.10_{-0.06}^{+0.07} $ &  $1.29_{-0.07}^{+0.06} $ &  $1.26_{-0.06}^{+0.07} $ \\
Stellar radius [\Rsun]  &  $1.39 \pm 0.11$ &  $1.40 \pm 0.10  $ &  $1.25 \pm 0.03 $  \\
Companion mass $M_{\rm b}$ [\Mjup ]  &  $18.0 \pm 0.9$ &  $20.1_{-1.2}^{+1.3} $ &  $19.1 \pm 1.0 $ \\
Companion radius $R_{\rm b}$ [\Rjup]  &  $1.22_{-0.10}^{+0.12}$ &  $1.24_{-0.10}^{+0.09} $ &  $1.11 \pm 0.03  $ \\
Companion density $\rho_{\rm b}$ [$\rm g\;cm^{-3}$] &  $12.4_{-2.6}^{+3.4}$ &  $13.0_{-2.2}^{+3.0} $ &  $17.4_{-1.4}^{+1.6} $ \\
Companion surface gravity log\,$g_{\rm b }$ [cgs] &  $4.48 \pm 0.09$ &  $4.51 \pm 0.05 $ &  $4.58 \pm 0.03 $ \\
Age $t$ [Gyr]  & $5.1 \pm 1.5$ &  $2.1_{-0.9}^{+0.8} $ &  $1.0_{-0.7}^{+0.9} $ \\
Orbital semi-major axis $a$ [au] & 0.155 $\pm$ 0.003 &  $0.164 \pm 0.003 $  & $0.162 \pm 0.003$ \\
Equilibrium temperature $T_{\rm eq}$ [K] ~$^b$& 905 $\pm$ 39 &  $897 \pm 29 $ &  $853 \pm 14 $\\
\hline       
\vspace{-0.5cm}
\footnotetext[1]{The limb-darkening coefficients were allowed to vary within their $1\sigma$ errors 
related to the uncertainties on stellar atmospheric parameters.} 
\footnotetext[2]{Black-body equilibrium temperature assuming a null Bond albedo and uniform 
heat redistribution to the night side.}
\end{tabular}
\end{minipage}
\label{table_param_KOI423}  
\end{table*}

\subsubsection{Kepler-41}
The new system parameters determined with SC data significantly
differ from those obtained by \citet{Santerneetal2011a}. The most
important difference in the fitted parameters is found for the orbital
inclination that, in turn, affects the determination of $a/R_\star$ and 
the stellar density. Indeed, while \citet{Santerneetal2011a} found $i=88.3 \pm 0.7$~deg,
our new solution points to a considerably higher impact parameter 
with  $i=82.51 \pm 0.09 $~deg (see Table~\ref{table_param_KOI196}).

By analysing only LC data, \citet{Southworth2012} found two solutions 
for the orbital inclination, one with $i \sim 90$~deg and the other one 
with $i \sim 80-82 $~deg. He opted for the first because 
``the $i \sim 80-82$~deg family occurs mainly for LD-fixed light-curve solutions,
and results in weird physical properties". However, by letting the LDC vary, 
our DE-MCMC run on SC data always converged toward the latter value
of $i$. This occurred even when the chains were all started at values close to $i \sim 90$~deg.
This ambivalence shows that, in some cases, fitting the LDC with LC data may 
lead to local minima that do not represent the true solution.

Long-cadence measurements were also used by \citet{Quintanaetal2013} 
to validate this planet by analysing the phase curve. Curiously,
these authors found an orbital inclination of $85.4_{-0.5}^{+0.4}$~deg, which is 
in between the other two solutions. However, we point out that 
\citet{Quintanaetal2013} were more interested in analysing the phase curve 
and did not explicitly mention which oversampling they adopted to 
model the LC transits.

Our solution with low orbital inclination $i=82.5$~deg 
is physically acceptable because the lowest possible
inclination for the transit of Kepler-41b is $\sim 78$~deg. 
The corresponding stellar density indicates a star that is larger 
and older than previously found by 
\citet{Santerneetal2011a} for the same atmospheric parameters.
The derived  $\log g=4.278 \pm 0.005 $ is now more consistent 
with the spectroscopic value $\log g=4.2 \pm 0.10 $ (see Table~\ref{table_param_KOI196}).
A larger stellar radius implies a larger planetary radius  $R_{\rm p}=1.29 \pm 0.02~\Rjup $, 
which now makes this planet more similar to the other close-in hot Jupiters. 
Indeed, from the planetary parameters that were previously determined 
by \citet{Santerneetal2011a} and \citet{Southworth2012}, 
this object appeared quite rare, meaning: non-inflated despite
its vicinity to the parent star (see Figs.~9 and 10 in \citealt{Santerneetal2011a}).

Figure~\ref{fig_KOI196} displays the phase-folded transit and radial-velocity 
measurements along with the best solution.

Last but not least, the lower inclination also has an impact on $a/\rp$
that becomes equal to $50.26 \pm 0.36$ and, in turn, 
affects the determination of the geometric albedo (cf., e.g., Eq.~(14) in \citealt{Roweetal2006}).
The latter was found to be considerably higher than
the majority of hot Jupiters with measured optical occultations 
by \citet{Santerneetal2011a}, that is $A_{\rm g}=0.30 \pm 0.07 $.
On the contrary, our new solution with SC data indicates 
a significantly lower geometric albedo of  $A_{\rm g}<0.135$
from the most recent value of the secondary-eclipse depth \citep{Angerhausenetal2014}, 
in agreement with theoretical expectations for atmospheres 
of hot Jupiters without scattering clouds (e.g., \citealt{Burrowsetal2008}).
Figure~\ref{fig_albedo_KOI196} shows the albedo values 
and their $1\sigma$ uncertainties as a function of the
planet day-side equilibrium temperature \unboldmath{$T_{\rm eq} $.
The vertical dashed lines
indicate the values of $T_{\rm eq}$ 
assuming perfect heat redistribution (left) 
or no redistribution in the atmosphere (right).

\begin{figure*}[t]
\centering
\begin{minipage}{15cm}
%\vspace{-0.5cm}
%\includegraphics[width=7cm]{plot_transit_KOI196_SC_2014.pdf}
\includegraphics[width=7cm]{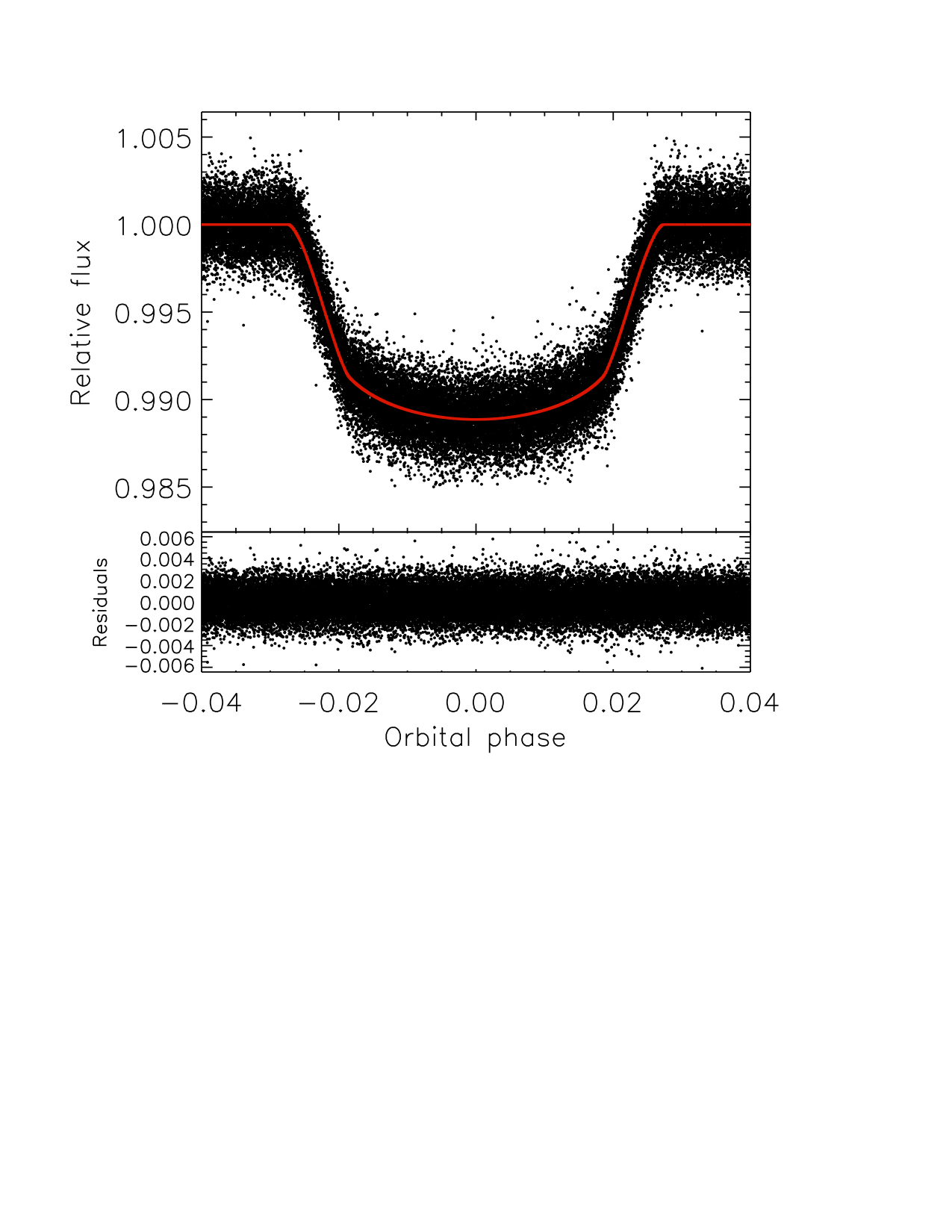}
\hspace{1.0cm}
\includegraphics[width=7cm]{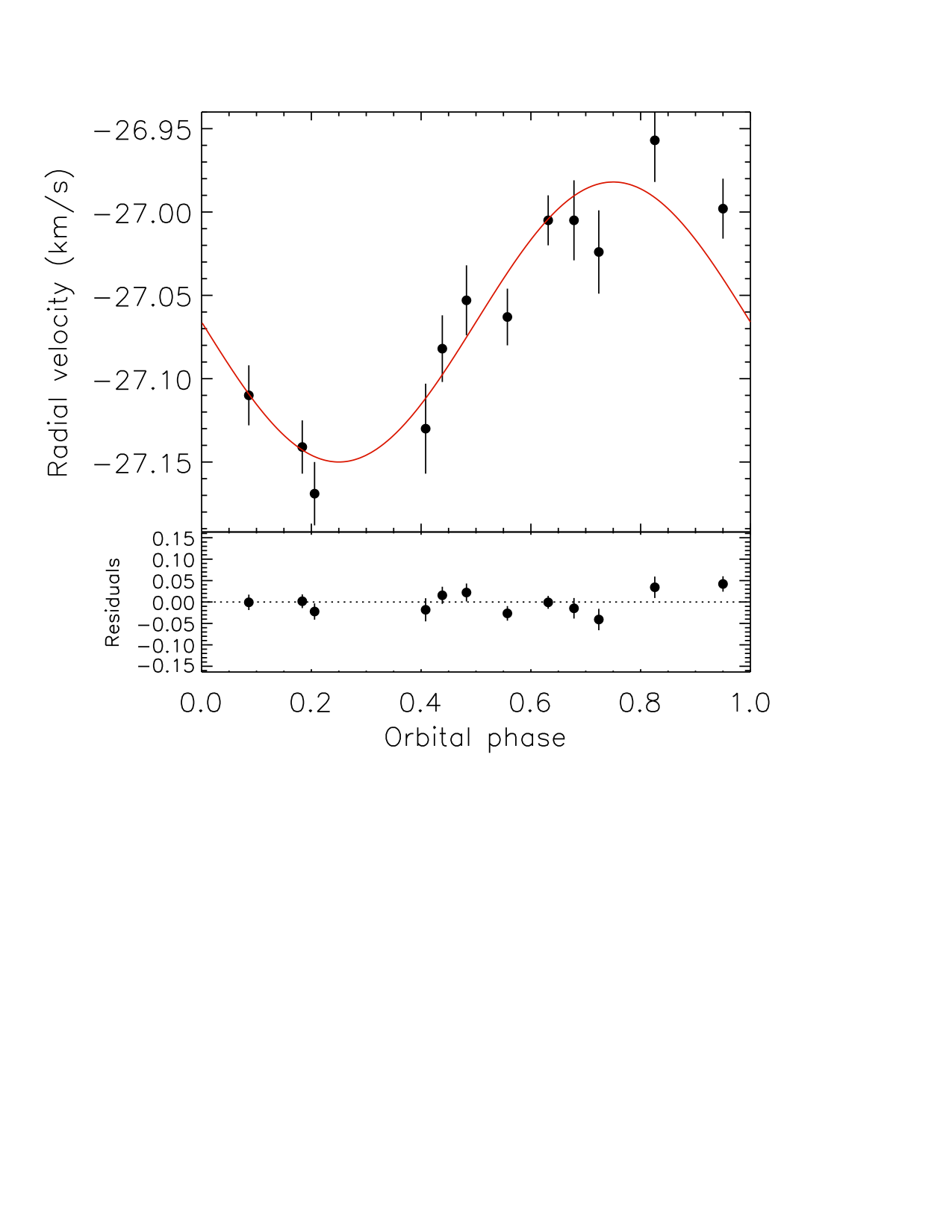}
\vspace{-3.7cm}
\caption{
\emph{Left panel}: phase-folded transit light curve of Kepler-41b along with the transit
model (red solid line).
\emph{Right panel}: phase-folded radial-velocity curve of Kepler-41 and, superimposed, the 
Keplerian model (red solid line).
}
\label{fig_KOI196}
\end{minipage}
\end{figure*}

\begin{table*}[h!]
\centering
\caption{Kepler-41 system parameters.}            
\begin{minipage}[t]{13.0cm} 
\setlength{\tabcolsep}{1.0mm}
\renewcommand{\footnoterule}{}                          
\begin{tabular}{l l l}        
\hline\hline                 
\emph{Fitted system parameters}  & \citet{Santerneetal2011a} & This work\\
\hline
Orbital period $P$ [days] & 1.855558 $\pm$ 0.000007 &  $1.85555820 \pm 0.00000052 $\\
Transit epoch $T_{ \rm 0} [\rm BJD_{TDB}-2454900$] & 70.1803 $\pm$ 0.0003 &  $287.280482 \pm 0.000051 $ \\
Transit duration $T_{\rm 14}$ [h] & 2.376 $\pm$ 0.048 &  $2.4341 \pm 0.0046 $ \\
Radius ratio $R_{\rm p}/R_{*}$ & $0.0895 \pm 0.0019$ &  $0.10265 \pm 0.00042 $  \\
Inclination $i$ [deg] & $88.3 \pm 0.7$ &  $82.51 \pm 0.09 $ \\
Limb-darkening coefficient $q_{1}$  &  - &  $0.39 \pm 0.04 $ \\
Limb-darkening coefficient $q_{2}$  &  - &  $0.37_{-0.07}^{+0.08} $  \\
Orbital eccentricity $e$  &  0 (fixed) & 0 (fixed)  \\
Radial velocity semi-amplitude $K$ [\ms] & $85 \pm 11$ &  $84 \pm 11 $ \\
Systemic velocity  $V_{\rm r}$ [\kms] & $-27.066 \pm 0.007$ & $-27.066 \pm 0.008$ \\
RV jitter [\ms] $s_{\rm j}$ & - &  $16_{-9}^{+10} $ \\
& \\
\multicolumn{2}{l}{\emph{Derived transit parameters}} \\
\hline
$a/R_{*}$ & $6.43 \pm 0.05$ &  $5.159 \pm 0.023 $ \\
Stellar density $\rho_{*}$ [$ \rm g\;cm^{-3}$] & $1.46 \pm 0.04 $ &  $0.754 \pm 0.010 $\\
Impact parameter $b$ & $0.19 \pm 0.07$ &  $0.672 \pm 0.005 $ \\
Limb-darkening coefficient $u_{a}$  & $ 0.57 \pm 0.13$&  $0.46 \pm 0.07 $ \\
Limb-darkening coefficient $u_{b}$  &  $0.41 \pm 0.13$ &  $0.16 \pm 0.10 $\\
& \\
\multicolumn{2}{l}{\emph{Atmospheric parameters of the star}} \\
\hline
Effective temperature $T_{\rm{eff}}$[K] & 5620 $\pm$ 140 &  $5750 \pm 100 $ \\
Spectroscopic surface gravity log\,$g$ [cgs]&  4.20  $\pm$ 0.15 &  $4.20  \pm 0.10 $ \\
Derived surface gravity log\,$g$ [cgs]&  4.47  $\pm$ 0.12 &  $4.278  \pm 0.005 $ \\
Metallicity $[\rm{Fe/H}]$ [dex] & 0.29  $\pm$ 0.16 &  $0.38 \pm 0.11 $ \\
Stellar rotational velocity $V \sin{i_{*}}$ [\kms] & 6 $\pm$ 2 & 6 $\pm$ 2 \\
Spectral type & G6V & G2V \\
& \\
\multicolumn{2}{l}{\emph{Stellar and planetary physical parameters}} \\
\hline
Stellar mass [\Msun] &  1.12 $\pm$ 0.07 &  $1.15 \pm 0.04 $ \\
Stellar radius [\Rsun] &  $1.02 \pm 0.03$ &  $1.29 \pm 0.02 $  \\
Planetary mass $M_{\rm p}$ [\Mjup ]  &  $0.55 \pm 0.09$ &  $0.56 \pm 0.08 $ \\
Planetary radius $R_{\rm p}$ [\Rjup]  &  $0.89 \pm 0.05$ &  $1.29 \pm 0.02 $ \\
Planetary density $\rho_{\rm p}$ [$\rm g\;cm^{-3}$] &  $1.10 \pm 0.18$ &  $0.33 \pm 0.04 $ \\
Planetary surface gravity log\,$g_{\rm p }$ [cgs] &  $3.23 \pm 0.09$ &  $2.92 \pm 0.06 $ \\
Age $t$ [Gyr]  & $0.6_{-0.3}^{+2.5}$ &  $4.4_{-1.1}^{+1.3} $ \\
Orbital semi-major axis $a$ [au] & 0.030 $\pm$ 0.010 &  $0.03101 \pm 0.0004 $ \\
Equilibrium temperature $T_{\rm eq}$ [K] ~$^a$ & 1730 $\pm$ 40 &  $1790 \pm 31 $\\
\hline
\vspace{-0.5cm}       
\footnotetext[1]{Black-body equilibrium temperature assuming a null Bond albedo and uniform 
heat redistribution to the night side.}
\end{tabular}
\end{minipage}
\label{table_param_KOI196}  
\end{table*}

\begin{figure}[]
\centering
\includegraphics[width=9cm]{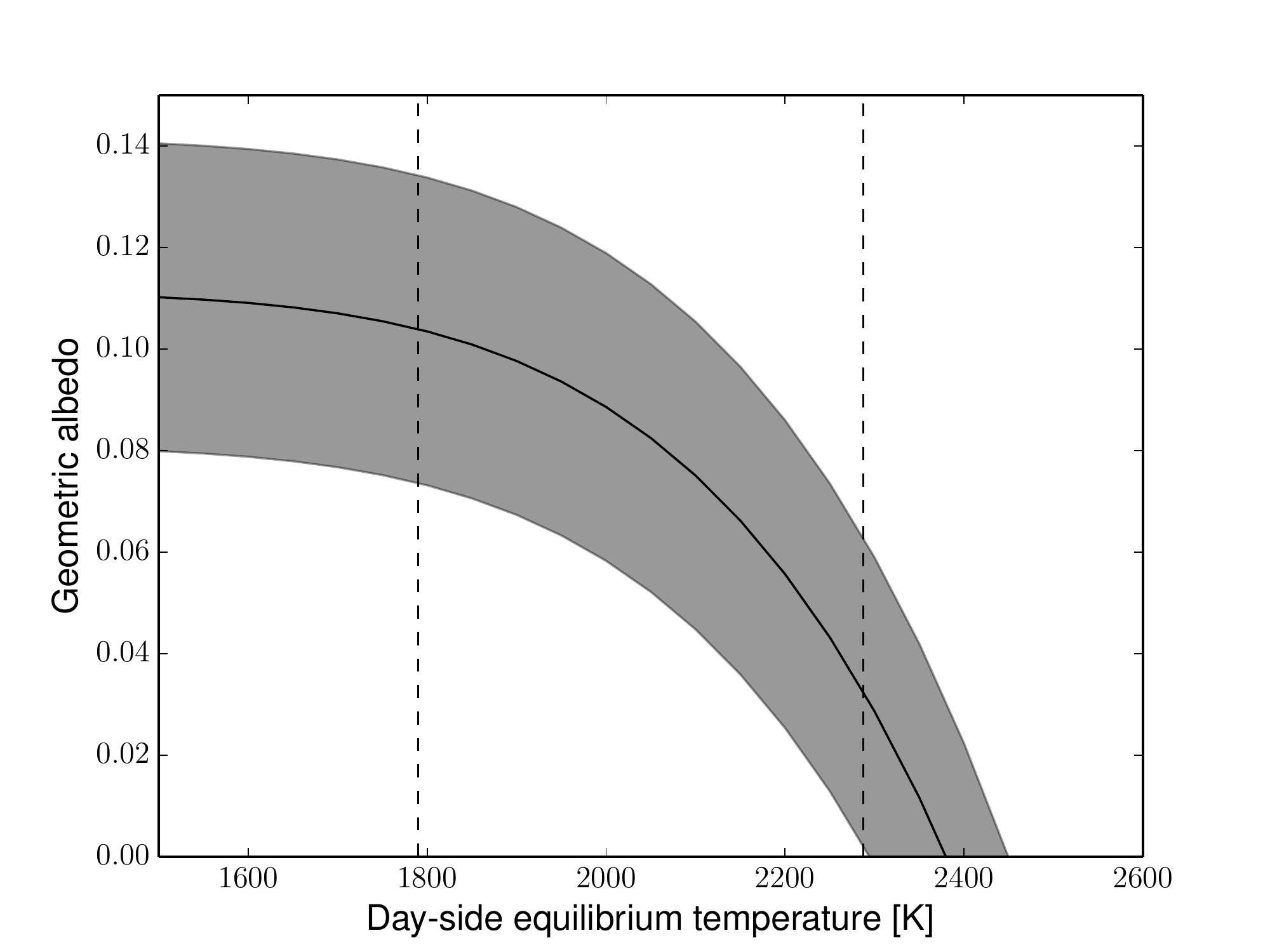}
\caption{Geometric albedo of Kepler-41b as a function of its day-side equilibrium temperature.
$T_{\rm eq}$. The vertical dashed lines indicate the values of $T_{\rm eq}$ 
assuming perfect heat redistribution (left) 
and no redistribution in the atmosphere (right). 
The grey band shows the albedo values
allowed by the $1\sigma$ uncertainty on the occultation depth 
determined by \citet{Santerneetal2011a}.
}
\label{fig_albedo_KOI196}
\end{figure}

\subsubsection{Kepler-43}

\noindent
System parameters derived from our DE-MCMC analysis are listed in
Table~\ref{table_param_KOI135}, and Fig.~\ref{fig_KOI135} shows
the phase-folded SC transit and radial-velocity curve along with
the best-fit model.
System parameters generally agree within 1~$\sigma$ 
with those determined by \citet{Bonomoetal2012a},
in spite of the very poor sampling of the phase-folded LC transit
(see their Fig.~5). The transit duration derived with SC data is 
slightly shorter (at $1.3~\sigma$) than found by \citet{Bonomoetal2012a}
but with negligible influence on system parameters.
The planet eccentricity is consistent with zero within $2~\sigma$.

From the stellar rotation period $P_{\rm rot}=12.95 \pm 0.25$~d derived with 
all the LC data, the gyrochronology age  $t_{\rm gyr}=1.7_{-0.4}^{+0.6} $~Gyr agrees well with 
that estimated from stellar evolutionary tracks  $2.3_{-0.7}^{+0.8} $~Gyr, as already noted by 
\citet{Bonomoetal2012a}.

\begin{figure*}[t]
\centering
\begin{minipage}{15cm}
%\vspace{-0.5cm}
%\includegraphics[width=7cm]{plot_transit_KOI135_SC_2014.pdf}
\includegraphics[width=7cm]{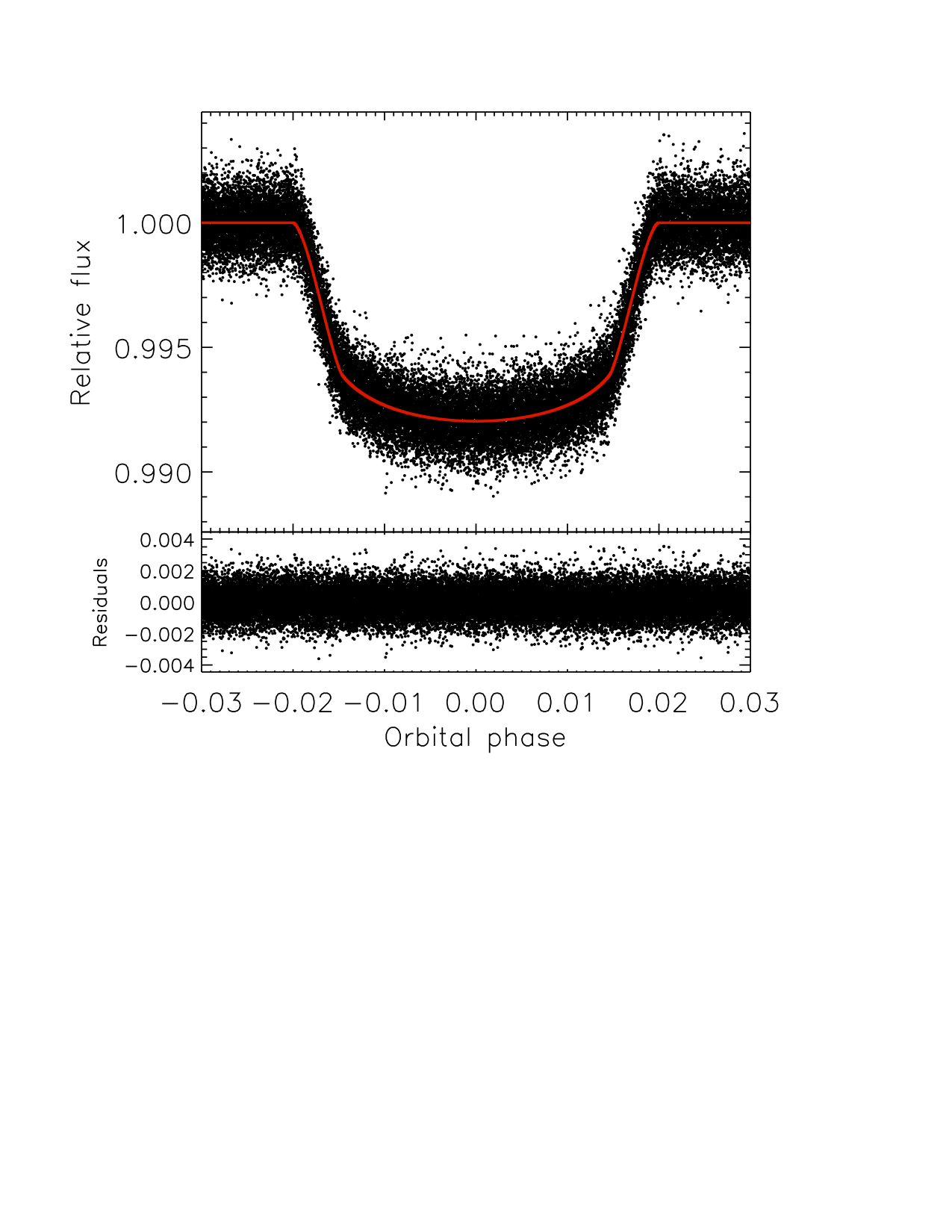}
\hspace{1.0cm}
\includegraphics[width=7cm]{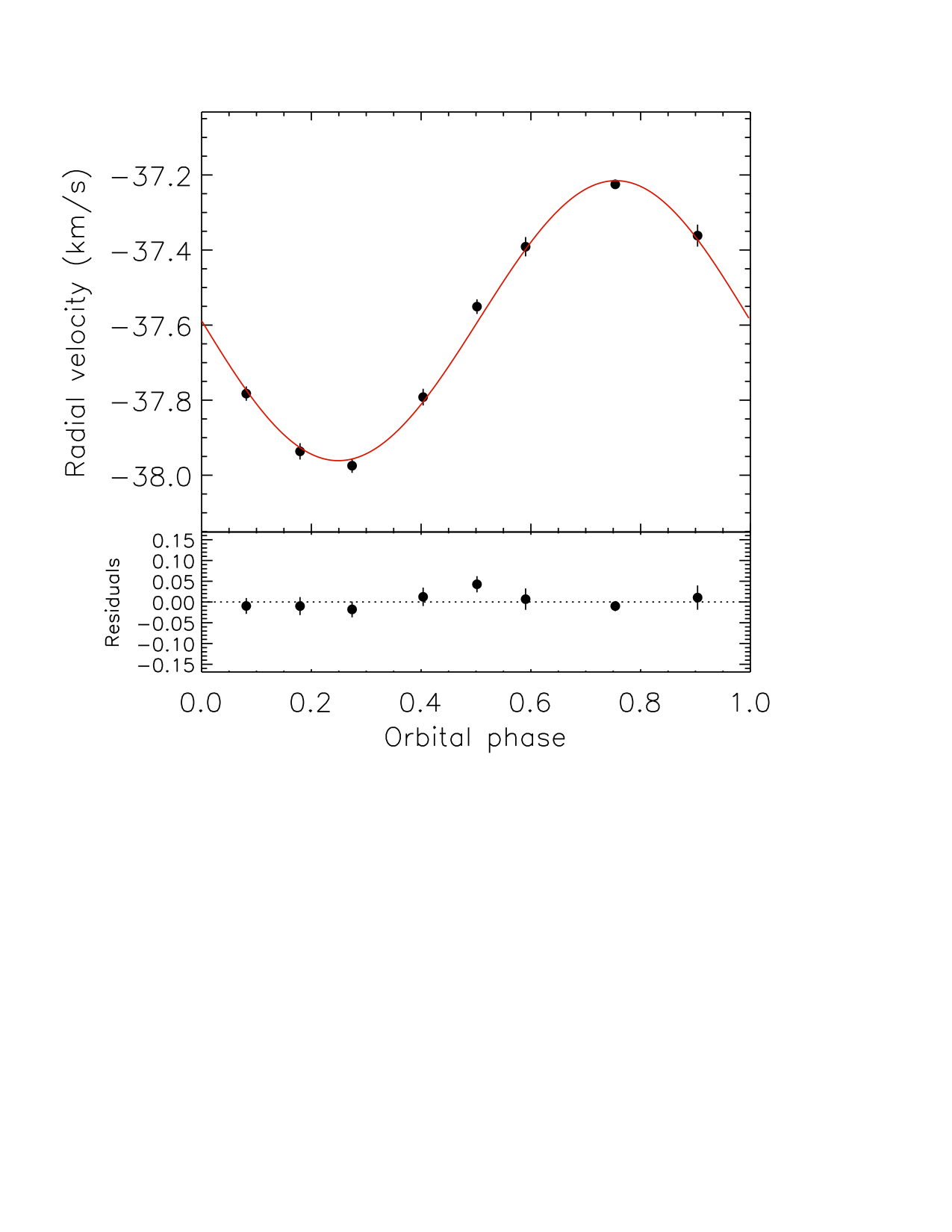}
\vspace{-3.7cm}
\caption{
\emph{Left panel}: phase-folded transit light curve of Kepler-43b along with the transit
model (red solid line).
\emph{Right panel}: phase-folded radial-velocity curve of Kepler-43 and, superimposed, the 
Keplerian model (red solid line).
}
\label{fig_KOI135}
\end{minipage}
\end{figure*}

\begin{table*}[h!]
\centering
\caption{Kepler-43 system parameters.}            
\begin{minipage}[t]{13.0cm} 
\setlength{\tabcolsep}{1.0mm}
\renewcommand{\footnoterule}{}                          
\begin{tabular}{l l l}        
\hline\hline                 
\emph{Fitted system parameters}  & \citet{Bonomoetal2012a} & This work\\
\hline
Orbital period $P$ [days] &  $3.024095 \pm 0.000021$ &  $3.02409309 \pm 0.00000020 $\\
Transit epoch $T_{ \rm 0} [\rm BJD_{TDB}-2454900$] & 165.4159 $\pm$ 0.0006 &  $195.45227 \pm 0.00005 $ \\
Transit duration $T_{\rm 14}$ [h] & 2.926 $\pm$ 0.019 &  $2.900 \pm 0.003 $ \\
Radius ratio $R_{\rm p}/R_{*}$ & $0.0868_{-0.0007}^{+0.0006}$ &  $0.08647 \pm 0.00019 $  \\
Inclination $i$ [deg] & $84.35_{-0.40}^{+0.47}$ &  $84.57_{-0.37}^{+0.18} $ \\
Limb-darkening coefficient $q_{1}$  & - &  $0.36 \pm 0.02 $ \\
Limb-darkening coefficient $q_{2}$   & - &  $0.30 \pm 0.04 $  \\
$\sqrt{e}~\cos{\omega}$ & - &  $0.061 \pm 0.030 $ \\
$\sqrt{e}~\sin{\omega}$  & - &  $0.086_{-0.14}^{+0.12} $ \\
Orbital eccentricity $e$  &  $< 0.025$ &  $0.017_{-0.009}^{+0.027} $ \\
Argument of periastron  [deg] $\omega$  &  - &  $52_{-81}^{+27} $  \\
Radial velocity semi-amplitude $K$ [\ms] & $375 \pm 13$ &  $373 \pm 9 $ \\
Systemic velocity  $V_{\rm r}$ [\kms] & $-37.591 \pm 0.007$ & $-37.591 \pm 0.007$ \\
RV jitter [\ms] $s_{\rm j}$ & - &  $< 6 $ \\
& \\
\multicolumn{2}{l}{\emph{Derived transit parameters}} \\
\hline
$a/R_{*}$ & $6.81_{-0.20}^{+0.24}$ &  $6.93_{-0.22}^{+0.11} $ \\
Stellar density $\rho_{*}$ [$ \rm g\;cm^{-3}$] & $0.65_{-0.05}^{+0.07}$ &  $0.69_{-0.06}^{+0.03} $\\
Impact parameter $b$ & $0.67_{-0.03}^{+0.02}$ &  $0.648 \pm 0.004 $ \\
Limb-darkening coefficient $u_{a}$  & $0.375 \pm 0.026$~$^a$ &  $0.36 \pm 0.04 $ \\
Limb-darkening coefficient $u_{b}$  & $0.277 \pm 0.015$~$^a$  &  $0.24 \pm 0.05 $\\
& \\
\multicolumn{2}{l}{\emph{Atmospheric parameters of the star}} \\
\hline
Effective temperature $T_{\rm{eff}}$[K] & 6041 $\pm$ 143 &  $6050 \pm 100 $ \\
Spectroscopic surface gravity log\,$g$ [cgs]&  4.64  $\pm$ 0.13 & 4.4  $\pm$ 0.1 \\
Derived surface gravity log\,$g$ [cgs]&  4.26  $\pm$ 0.05 &  $4.26  \pm 0.02 $ \\
Metallicity $[\rm{Fe/H}]$ [dex] & 0.33  $\pm$ 0.11 &  $0.40  \pm 0.10 $ \\
Stellar rotational velocity $V \sin{i_{*}}$ [\kms] & 5.5 $\pm$ 1.5 & 5.5 $\pm$ 1.5 \\
Spectral type & F8V & F8V \\
& \\
\multicolumn{2}{l}{\emph{Stellar and planetary physical parameters}} \\
\hline
Stellar mass [\Msun] &  1.32 $\pm$ 0.09 &  $1.27 \pm 0.04 $ \\
Stellar radius [\Rsun] &  $1.42 \pm 0.07$ &  $1.38_{-0.03}^{+0.05} $  \\
Planetary mass $M_{\rm p}$ [\Mjup ]  &  $3.23 \pm 0.19$ &  $3.13 \pm 0.10 $ \\
Planetary radius $R_{\rm p}$ [\Rjup]  &  $1.20 \pm 0.06$ &  $1.16_{-0.03}^{+0.04} $ \\
Planetary density $\rho_{\rm p}$ [$\rm g\;cm^{-3}$] &  $2.33 \pm 0.36$ &  $2.49_{-0.23}^{+0.16} $ \\
Planetary surface gravity log\,$g_{\rm p }$ [cgs] &  $3.75 \pm 0.04$ &  $3.76_{-0.03}^{+0.02} $ \\
Age $t$ [Gyr]  & $2.8^{+1.0}_{-0.8}$ &  $2.3_{-0.7}^{+0.8} $ \\
Orbital semi-major axis $a$ [au] & 0.0449 $\pm$ 0.0010 &  $0.0444 \pm 0.0005 $ \\
Equilibrium temperature $T_{\rm eq}$ [K] ~$^b$ & 1637 $\pm$ 47 &  $1628 \pm 33 $\\
\hline       
\vspace{-0.5cm}
\footnotetext[1]{The limb-darkening coefficients were allowed to vary within their $1\sigma$ errors 
related to the uncertainties on stellar atmospheric parameters.} 
\footnotetext[2]{Black-body equilibrium temperature assuming a null Bond albedo and uniform 
heat redistribution to the night side.}
\end{tabular}
\end{minipage} 
\label{table_param_KOI135}  
\end{table*}

\subsubsection{Kepler-44}
New system parameters slightly differ from those derived by \citet{Bonomoetal2012a} 
(see Table~\ref{table_param_KOI204}). 
Our analysis with SC data reveals a transit duration that is $2\sigma$ shorter than 
found by \citet{Bonomoetal2012a}. This implies a slightly larger $a/R_\star$,
hence a moderately higher stellar density from the Kepler third law.
In consequence, stellar evolutionary tracks point to 
a slightly smaller star than previously reported by \citet{Bonomoetal2012a}, with 
radius and mass of $1.35 \pm 0.08~\Rsun$ and
$1.12 \pm 0.08~\Msun$, respectively.
Therefore, also the planet turns out to be
smaller (from $\rp/R_\star$) and denser:
$R_{\rm p}=1.09 \pm 0.07~\Rjup$, $M_{\rm p}=1.00 \pm 0.10~\Mjup$, and 
$\rho_{\rm p}=0.93_{-0.17}^{+0.19}~\rm g\;cm^{-3}$
(see Table~\ref{table_param_KOI204}).

Figure~\ref{fig_KOI204} shows the phase-folded transit and radial-velocity 
curve and, superimposed, the transit and the Keplerian models.

\begin{figure*}[t]
\centering
\begin{minipage}{15cm}
%\vspace{-0.5cm}
%\includegraphics[width=7cm]{plot_transit_KOI204_SC_2014.pdf}
\includegraphics[width=7cm]{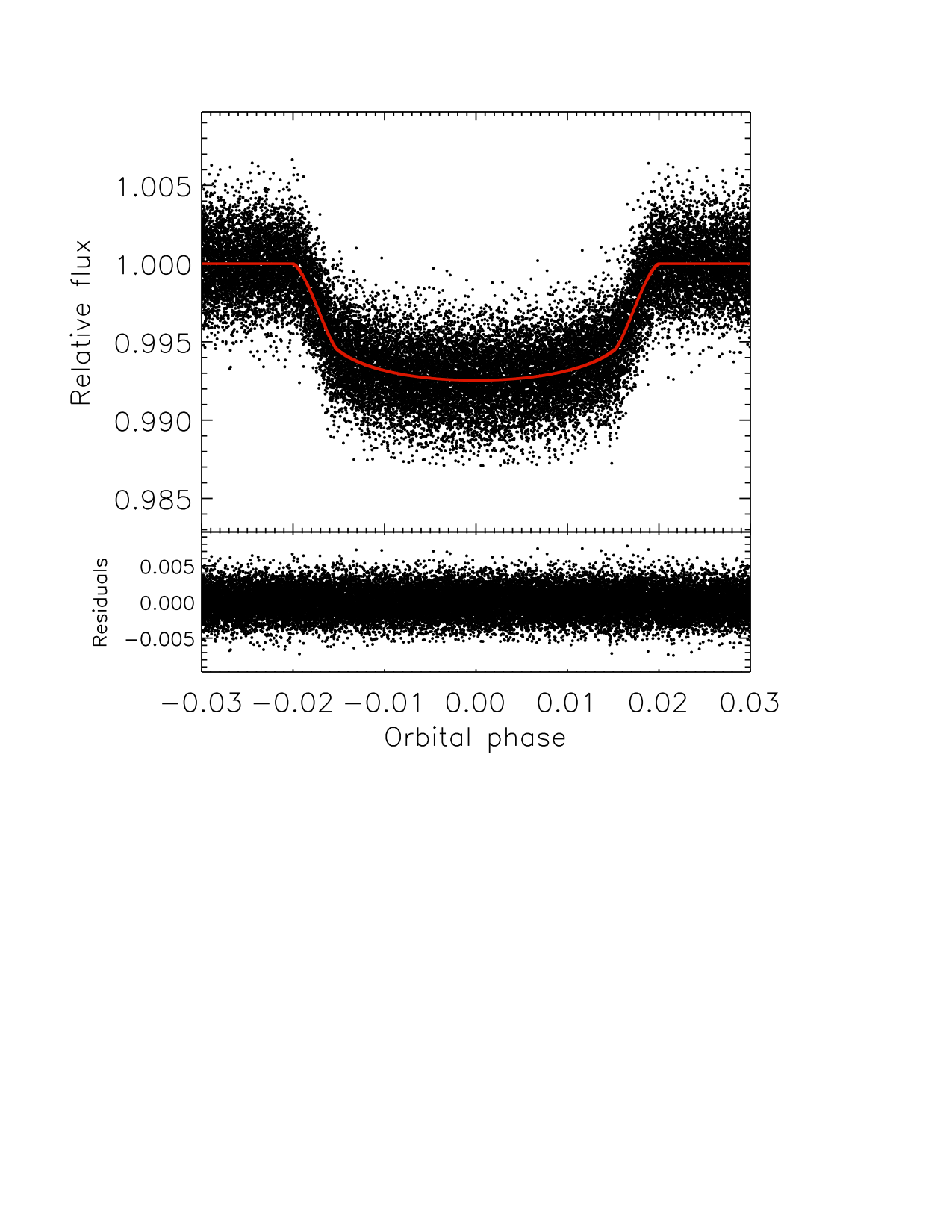}
\hspace{1.0cm}
\includegraphics[width=7cm]{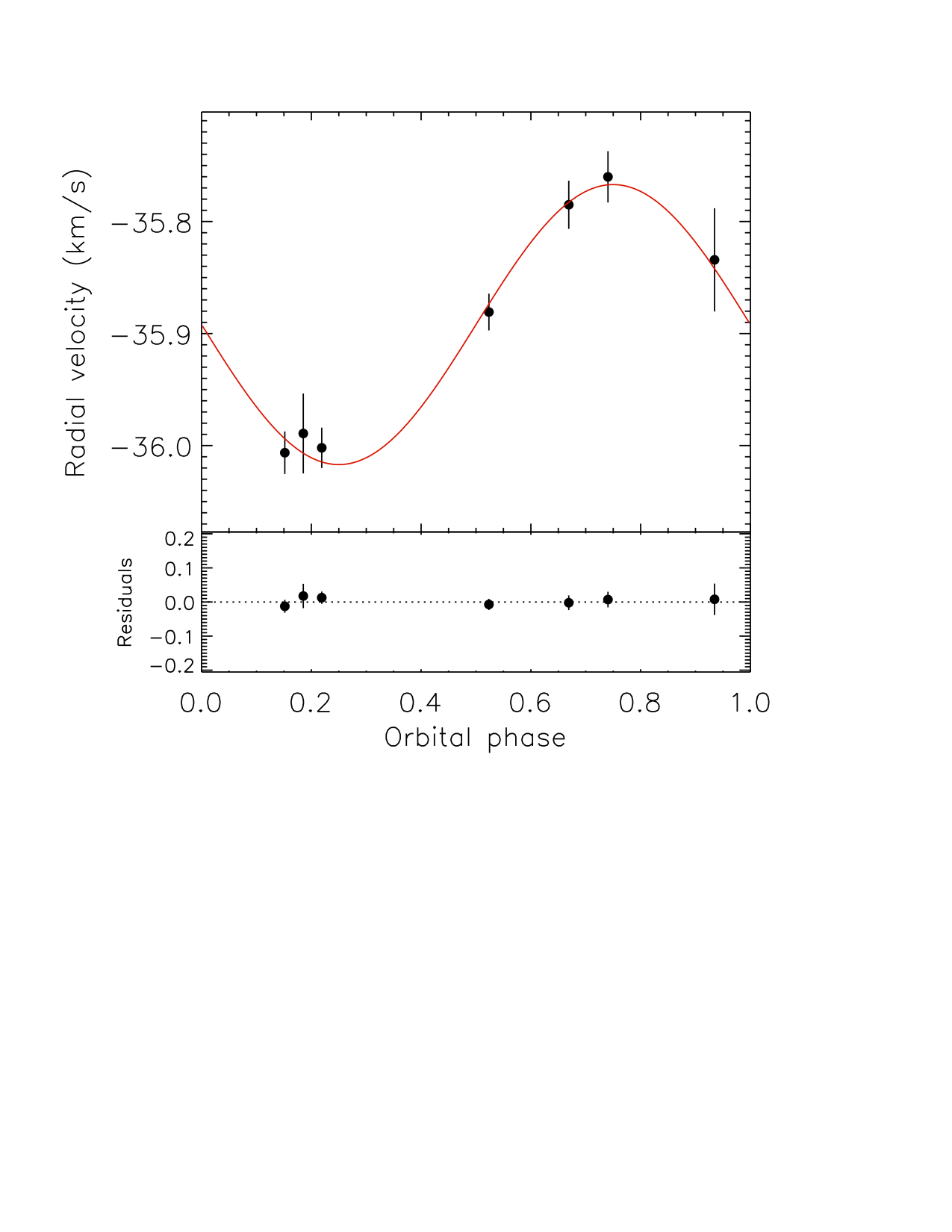}
\vspace{-3.7cm}
\caption{
\emph{Left panel}: phase-folded transit light curve of Kepler-44b along with the transit
model (red solid line).
\emph{Right panel}: phase-folded radial-velocity curve of Kepler-44 and, superimposed, the 
Keplerian model (red solid line).
}
\label{fig_KOI204}
\end{minipage}
\end{figure*}

\begin{table*}[h!]
\centering
\caption{Kepler-44 system parameters.}            
\begin{minipage}[t]{13.0cm} 
\setlength{\tabcolsep}{1.0mm}
\renewcommand{\footnoterule}{}                          
\begin{tabular}{l l l}        
\hline\hline                 
\emph{Fitted system parameters}  & \citet{Bonomoetal2012a} & This work\\
\hline
Orbital period $P$ [days] & 3.246740 $\pm$ 0.000018 &  $3.2467293 \pm 0.0000030 $ \\
Transit epoch $T_{ \rm 0} [\rm BJD_{TDB}-2454900$] & 166.3781 $\pm$ 0.0004 &  $287.15640 \pm 0.00021 $ \\
Transit duration $T_{\rm 14}$ [h] & 3.218 $\pm$ 0.043 &  $3.124 \pm 0.014 $ \\
Radius ratio $R_{\rm p}/R_{*}$ & 0.0844 $\pm$ 0.0011 &  $0.0828 \pm 0.0008 $  \\
Inclination $i$ [deg] & $83.78_{-0.55}^{+0.65}$ &  $84.96_{-0.62}^{+0.50} $ \\
Limb-darkening coefficient $q_{1}$  & - &  $ 0.40_{-0.08}^{+0.10}  $ \\
Limb-darkening coefficient $q_{2}$  & - &  $ 0.33_{-0.13}^{+0.15}  $  \\
$\sqrt{e}~\cos{\omega}$ & - &  $ 0.05_{-0.15}^{+0.12}  $ \\
$\sqrt{e}~\sin{\omega}$  & - &  $ 0.03 \pm 0.20  $ \\
Orbital eccentricity $e$  &  $< 0.021$ &  $ < 0.066 $  \\
Radial velocity semi-amplitude $K$ [\ms] & $124 \pm 5$ & $125 \pm 11$ \\
Systemic velocity  $V_{\rm r}$ [\kms] & $-35.892 \pm 0.004$ & $-35.892 \pm 0.009$ \\
RV jitter [\ms] $s_{\rm j}$ & - &  $< 7 $ \\
& \\
\multicolumn{2}{l}{\emph{Derived transit parameters}} \\
\hline
$a/R_{*}$ & $6.45_{-0.26}^{+0.32}$ &  $7.07_{-0.37}^{+0.35} $ \\
Stellar density $\rho_{*}$ [$ \rm g\;cm^{-3}$] & $0.48_{-0.06}^{+0.07}$ &  $0.63 \pm 0.10 $ \\
Impact parameter $b$ & $0.70_{-0.04}^{+0.03}$ &  $0.62 \pm 0.02 $ \\
Limb-darkening coefficient $u_{a}$  & $0.420 \pm 0.026$~$^a$ &  $0.42 \pm 0.13 $ \\
Limb-darkening coefficient $u_{b}$  & $0.248 \pm 0.017$~$^a$  &  $0.21 \pm 0.19 $ \\
& \\
\multicolumn{2}{l}{\emph{Atmospheric parameters of the star}} \\
\hline
Effective temperature $T_{\rm{eff}}$[K] & 5757 $\pm$ 134 &  $5800 \pm 100 $ \\
Metallicity $[\rm{Fe/H}]$ [dex] & 0.26  $\pm$ 0.10 &  $0.15  \pm 0.10 $ \\
Spectroscopic surface gravity log\,$g$ [cgs]&  4.59  $\pm$ 0.14 & 4.1  $\pm$ 0.1 \\
Derived surface gravity log\,$g$ [cgs]&  4.15  $\pm$ 0.06 &  $4.22  \pm 0.04 $ \\
Stellar rotational velocity $V \sin{i_{*}}$ [\kms] & 4 $\pm$ 2 & 4 $\pm$ 2 \\
Spectral type & G2IV & G2IV \\
& \\
\multicolumn{2}{l}{\emph{Stellar and planetary physical parameters}} \\
\hline
Stellar mass [\Msun] &  $1.19 \pm 0.10$ &  $1.12 \pm 0.08 $ \\
Stellar radius [\Rsun] &  $1.52 \pm 0.09$ &  $1.35 \pm 0.08 $  \\
Planetary mass $M_{\rm p}$ [\Mjup ]  &  $1.02 \pm 0.07$ & $1.00 \pm 0.10$ \\
Planetary radius $R_{\rm p}$ [\Rjup]  &  $1.24 \pm 0.07$ &  $1.09 \pm 0.07 $ \\
Planetary density $\rho_{\rm p}$ [$\rm g\;cm^{-3}$] &  $0.65 \pm 0.12$ &  $0.93_{-0.17}^{+0.19} $ \\
Planetary surface gravity log\,$g_{\rm p }$ [cgs] &  $3.21 \pm 0.05$ & $3.31 \pm 0.06$ \\
Age $t$ [Gyr]  & $6.95^{+1.1}_{-1.7}$ &  $5.8_{-1.5}^{+2.4} $ \\
Orbital semi-major axis $a$ [au] &  0.0455 $\pm$ 0.0013 &  $0.0446 \pm 0.0011 $ \\
Equilibrium temperature $T_{\rm eq}$ [K] ~$^b$ & 1603 $\pm$ 51 &  $1544 \pm 47 $\\
\hline       
\vspace{-0.5cm}
\footnotetext[1]{The limb-darkening coefficients were allowed to vary within their $1\sigma$ errors 
related to the uncertainties on stellar atmospheric parameters.} 
\footnotetext[2]{Black-body equilibrium temperature assuming a null Bond albedo and uniform 
heat redistribution to the night side.}
\end{tabular}
\end{minipage}
\label{table_param_KOI204}  
\end{table*}

\subsubsection{Kepler-74}
The circular solution we decided to adopt differs from the eccentric system parameters by 
almost three standard deviations. This is mainly because the transit 
density used to determine stellar parameters is a function of the eccentricity.
For a null eccentricity, the planet becomes smaller and denser, with
$R_{\rm p}=0.97 \pm 0.04~\Rjup$, $M_{\rm p}=0.64 \pm 0.10~\Mjup$, and 
$\rho_{\rm p}=0.86 \pm 0.18~\rm g\;cm^{-3}$. 
The best fit of the transit and radial velocities is displayed in Fig.~\ref{fig_KOI200}.

\begin{figure*}[t]
\centering
\begin{minipage}{15cm}
%\vspace{-0.5cm}
%\includegraphics[width=7.0cm]{plot_transit_KOI200_SC_2014.pdf}
\includegraphics[width=7.0cm]{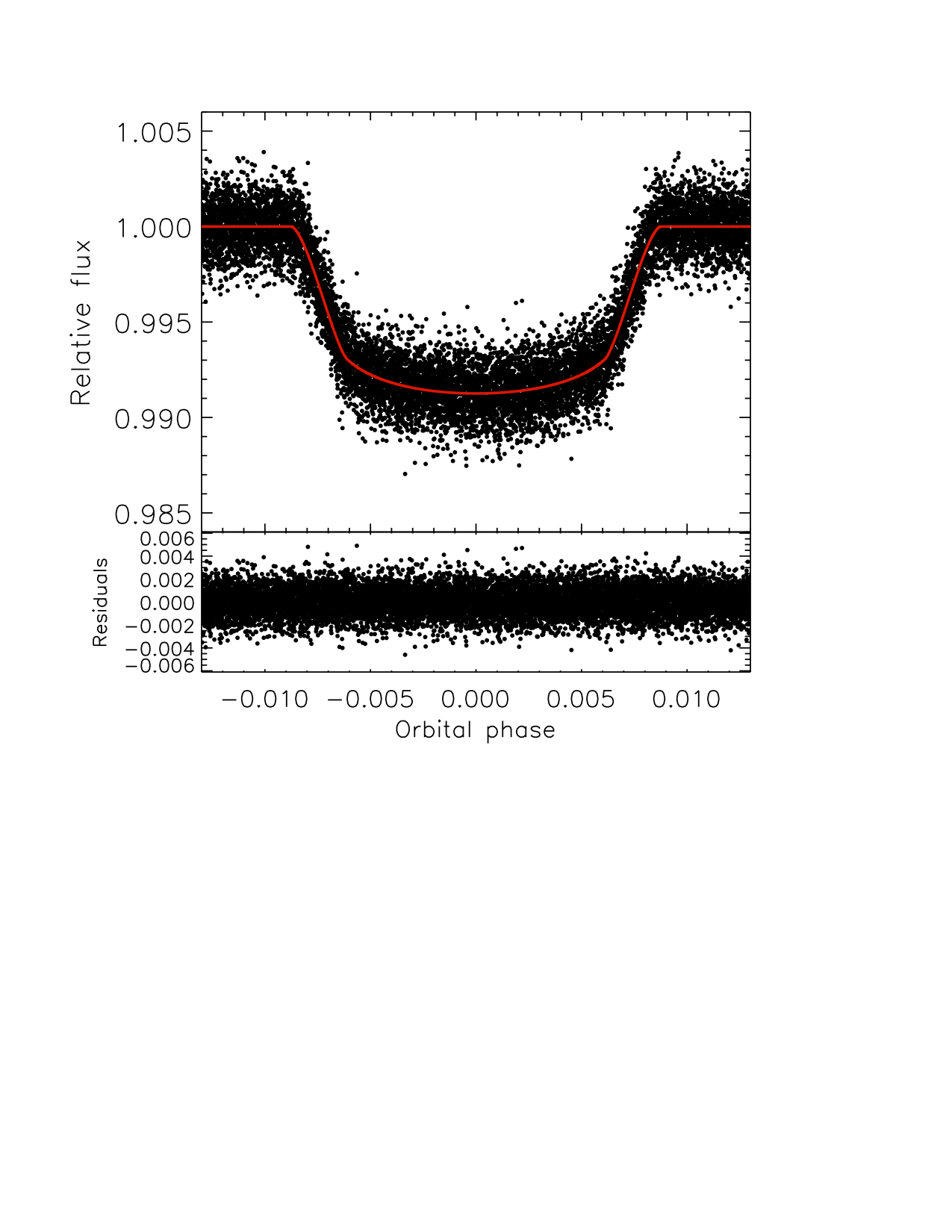}
\hspace{1.0cm}
\includegraphics[width=7.0cm]{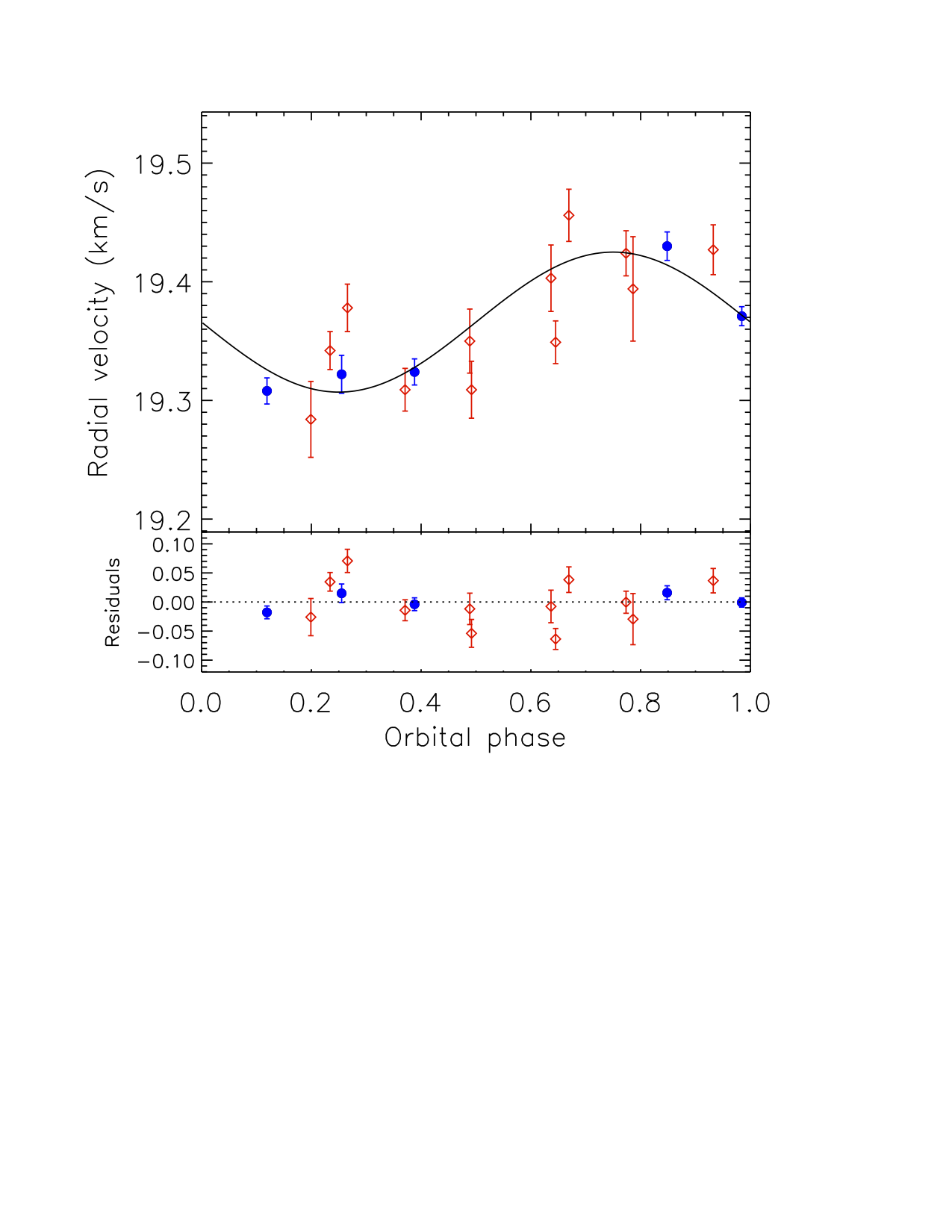}
\vspace{-3.7cm}
\caption{
\emph{Left panel}: phase-folded transit light curve of Kepler-74b along with the transit
model for a circular orbit and their residuals. 
\emph{Right panel}: phase-folded radial-velocity curve of Kepler-74. Red diamonds and blue
circles show SOPHIE and HARPS-N radial velocities, respectively. The black solid line
displays the Keplerian circular model.
}
\label{fig_KOI200}
\end{minipage}
\end{figure*}

\begin{table*}[h!]
\centering
\caption{Kepler-74 system parameters.}            
\begin{minipage}[t]{15.0cm} 
\setlength{\tabcolsep}{1.2mm}
\renewcommand{\footnoterule}{}                          
\begin{tabular}{l l l}        
\hline\hline                 
\emph{Fitted system parameters}  & \citet{Hebrardetal2013} &  This work \\
\hline
Orbital period $P$ [days] & 7.340718 $\pm$ 0.000001 &  $7.340711 \pm 0.000006 $  \\
Transit epoch $T_{ \rm 0} [\rm BJD_{TDB}-2454900$] & 67.3453 $\pm$ 0.0003 &  $287.56737 \pm 0.00014 $ \\
Transit duration $T_{\rm 14}$ [h] & 2.699 $\pm$ 0.038~\footnote{This value 
of transit duration is not accurate enough because it 
was estimated with the approximated formula reported in \citet{Winn2010}. Indeed, the transit duration 
is not a free parameter in the transit model adopted by \citet{Hebrardetal2013}. } 
&  $3.082 \pm 0.011 $ \\
Radius ratio $R_{\rm p}/R_{*}$ & $0.090 \pm 0.002$ &  $0.0912 \pm 0.0009 $  \\
Inclination $i$ [deg] & $85.55 \pm 0.96$ &  $87.46 \pm 0.07 $ \\
Limb-darkening coefficient $q_{1}$  & - &  $0.36_{-0.07}^{+0.10} $ \\
Limb-darkening coefficient $q_{2}$  & - &   $0.29_{-0.16}^{+0.19} $  \\
Orbital eccentricity $e$  &  $0.287 \pm 0.062$ & 0 (fixed)  \\
Argument of periastron $\omega$ [deg] & 64 $\pm$ 21 & -\\ 
Radial velocity semi-amplitude $K$ [\ms] & $58 \pm 7$ &  $59 \pm 11 $ \\
HARPS-N systemic velocity  $V_{\rm r, HN}$ [\kms] & $19.356 \pm 0.008$ &   $19.366 \pm 0.009 $ \\
HARPS-N RV jitter [\ms] $s_{\rm j, HN}$ &  $ < 5 $ &  $12_{-9}^{+16} $ \\
SOPHIE systemic velocity  $V_{\rm r, SO}$ [\kms] & $19.293_{-0.014}^{+0.008}$ &  $19.290 \pm 0.014 $ \\
SOPHIE RV jitter [\ms] $s_{\rm j, SO}$ &  $ 28 \pm 13 $ &  $40_{-10}^{+15} $ \\
& \\
\multicolumn{2}{l}{\emph{Derived transit parameters}} \\
\hline
$a/R_{*}$ & $11.8_{-0.8}^{+1.4}$ &  $15.47 \pm 0.18 $ \\
Stellar density $\rho_{*}$ [$ \rm g\;cm^{-3}$] & $0.58_{-0.11}^{+0.22}$ &  $1.30_{-0.04}^{+0.05} $ \\
Impact parameter $b$ & $0.684 \pm 0.032$ &  $0.685 \pm 0.011 $ \\
Limb-darkening coefficient $u_{a}$  & $0.10_{-0.17}^{+0.25}$ &  $0.35 \pm 0.17 $ \\
Limb-darkening coefficient $u_{b}$  &  $0.6 \pm 0.4$  &  $0.25_{-0.23}^{+0.25} $ \\
& \\
\multicolumn{2}{l}{\emph{Atmospheric parameters of the star}} \\
\hline
Effective temperature $T_{\rm{eff}}$[K] & 6050 $\pm$ 110 &  $6000 \pm 100 $ \\
Spectroscopic surface gravity log\,$g$ [cgs]& $4.2 \pm 0.1$ & 4.5  $\pm$ 0.10 \\
Derived surface gravity log\,$g$ [cgs]&  4.2  $\pm$ 0.1 &  $4.44 \pm 0.01 $ \\
Metallicity $[\rm{Fe/H}]$ [dex] & 0.34  $\pm$ 0.14 &  $0.42 \pm 0.11 $ \\
Stellar rotational velocity $V \sin{i_{*}}$ [\kms] & 5.0 $\pm$ 1.0 & 5.0 $\pm$ 1.0 \\
Spectral type & F8V & F8V \\
& \\
\multicolumn{2}{l}{\empty{Stellar and planetary physical parameters}} \\
\hline
Stellar mass [\Msun] &  $ 1.40_{-0.11}^{+0.14} $ &  $1.18 \pm 0.04 $ \\
Stellar radius [\Rsun] &  $1.51 \pm 0.14$ &  $1.12 \pm 0.04 $  \\
Planetary mass $M_{\rm p}$ [\Mjup ]  &  $0.68 \pm 0.09$ &  $0.63 \pm 0.12 $ \\
Planetary radius $R_{\rm p}$ [\Rjup]  &  $1.32 \pm 0.14$ &  $0.96 \pm 0.02  $ \\
Planetary density $\rho_{\rm p}$ [$\rm g\;cm^{-3}$] &  $0.37 \pm 0.13$ &  $0.88 \pm 0.18 $ \\
Planetary surface gravity log\,$g_{\rm p }$ [cgs] &  $ 2.98 \pm 0.11$ &  $3.23 \pm 0.10 $ \\
Age $t$ [Gyr]  & $2.9^{+1.5}_{-0.8}$ &  $0.8_{-0.5}^{+0.9} $ \\
Orbital semi-major axis $a$ [au] & 0.084 $\pm$ 0.014 &  $0.0781 \pm 0.0007 $ \\
Equilibrium temperature $T_{\rm eq}$ [K] ~$^b$ & 1250 $\pm$ 120 &  $1078 \pm 19 $ \\
\hline
\vspace{-0.5 cm}
\footnotetext[2]{Black-body equilibrium temperature assuming a null Bond albedo and uniform 
heat redistribution to the night side.}       
\end{tabular}
\end{minipage}
\label{table_param_KOI200}  
\end{table*}

\subsubsection{Kepler-75}
The agreement between the system parameters 
determined by \citet{Hebrardetal2013}
and our DE-MCMC solution obtained with SC data
(see Table~\ref{table_param_KOI889}) is excellent, even 
adopting the slightly different atmospheric parameters derived with 
MOOG: $T_{\rm eff}=5200 \pm 100$~K and $\rm [Fe/H]=0.30 \pm 0.12$. 

The stellar rotation period inferred from the whole LC light curve
is $P_{\rm rot} = 19.18 \pm 0.15$~d, in agreement with \citet{Hebrardetal2013}.
The system age estimated from gyrochronology 
\citep{MamajekHillenbrand2008} is $1.6 \pm 0.3$~Gyr, 
which is slightly lower than the value provided by stellar models, although
consistent with the latter at $1.7~\sigma$.

The transit and radial velocity data along with the best solutions 
are shown in Fig.~\ref{fig_KOI889}.

\begin{figure*}[t]
\centering
\begin{minipage}{15cm}
%\vspace{-0.5cm}
%\includegraphics[width=7cm]{plot_transit_KOI889_SC_2014.pdf}
\includegraphics[width=7cm]{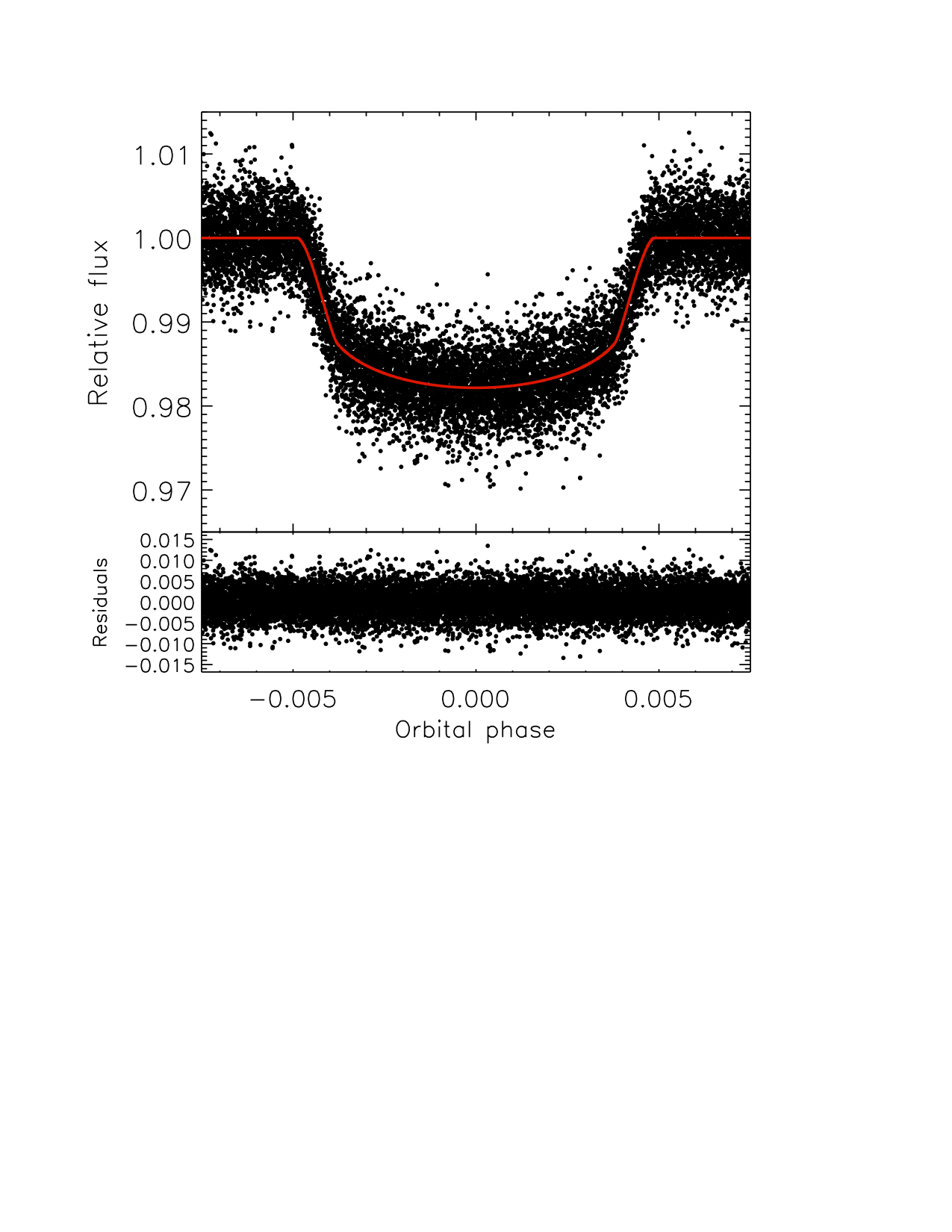}
\hspace{1.0cm}
\includegraphics[width=7cm]{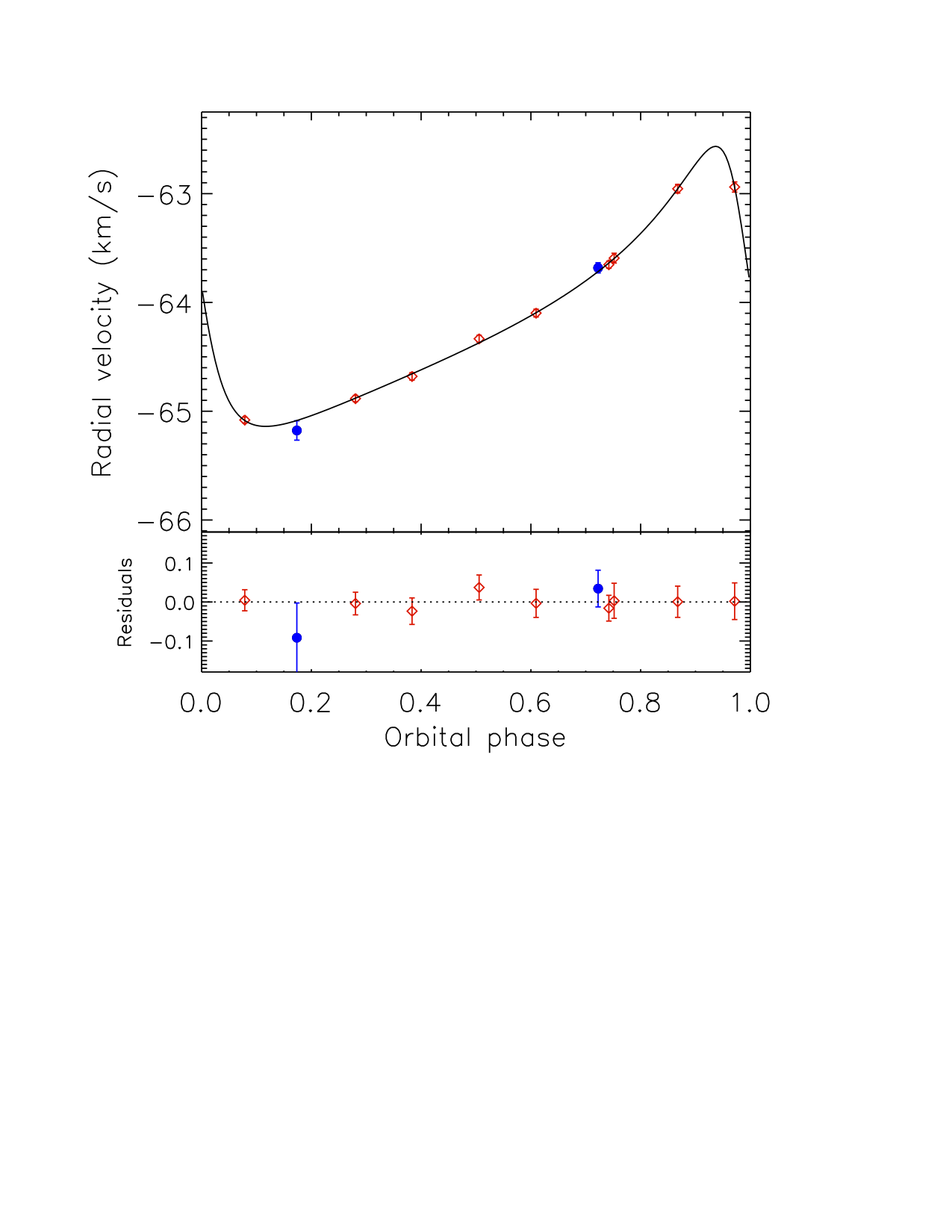}
\vspace{-3.7cm}
\caption{
\emph{Left panel}: phase-folded transit light curve of Kepler-75b along with the transit
model (red solid line).
\emph{Right panel}: phase-folded radial-velocity curve of Kepler-75. Red diamonds and blue
circles show SOPHIE and HARPS-N radial velocities, respectively. The black solid line
displays the Keplerian model.
}
\label{fig_KOI889}
\end{minipage}
\end{figure*}

\begin{table*}[h!]
\centering
\caption{Kepler-75 system parameters.}            
\begin{minipage}[t]{13.0cm} 
\setlength{\tabcolsep}{1.0mm}
\renewcommand{\footnoterule}{}                          
\begin{tabular}{l l l}        
\hline\hline                 
\emph{Fitted system parameters}  & \citet{Hebrardetal2013} & This work \\
\hline
Orbital period $P$ [days] & 8.884924 $\pm$ 0.000002 &  $8.8849116 \pm 0.0000034 $ \\
Transit epoch $T_{ \rm 0} [\rm BJD_{TDB}-2454900$] & 102.9910 $\pm$ 0.0002 &  $840.44030 \pm 0.00012 $ \\
Transit duration $T_{\rm 14}$ [h] & 1.872 $\pm$ 0.025 ~\footnote{This value 
of transit duration is not accurate enough because it 
was estimated with the approximated formula reported in \citet{Winn2010}. Indeed, the transit duration 
is not a free parameter in the transit model adopted by \citet{Hebrardetal2013}. } &  $2.0778 \pm 0.0084 $ \\
Radius ratio $R_{\rm p}/R_{*}$ & $0.121 \pm 0.002$ &  $0.1212_{-0.0007}^{+0.0009} $  \\
Inclination $i$ [deg] & $89.1_{-1.0}^{+0.6} $ &  $89.12_{-0.64}^{+0.51} $ \\
Limb-darkening coefficient $q_{1}$  & - &  $0.36 \pm 0.08 $ \\
Limb-darkening coefficient $q_{2}$  & - &  $0.42_{-0.08}^{+0.10} $  \\
$\sqrt{e}~\cos{\omega}$ & &  $0.337 \pm 0.020 $ \\
$\sqrt{e}~\sin{\omega}$  & &  $0.676 \pm 0.013 $ \\
Orbital eccentricity $e$  &  0.569 $\pm$ 0.010 &  $0.570 \pm 0.010 $ \\
Argument of periastron $\omega$ [deg] & 63.6 $\pm$ 1.4 &  63.5 $\pm$ 1.7\\ 
Radial velocity semi-amplitude $K$ [\kms] & $1.288 \pm 0.024$ &  $1.287 \pm 0.025 $ \\
HARPS-N systemic velocity  $V_{\rm r, HN}$ [\kms] & $-64.175 \pm 0.050$ &  $-64.180 \pm 0.056 $ \\
HARPS-N RV jitter [\ms] $s_{\rm j, HN}$ &  $ < 17 $ &  $< 54 $ \\
SOPHIE systemic velocity  $V_{\rm r, SO}$ [\kms] & $-64.235 \pm 0.012$ &  $-64.236 \pm 0.014 $ \\
SOPHIE RV jitter [\ms] $s_{\rm j, SO}$ &  $ < 8 $ &  $< 8 $ \\
& \\
\multicolumn{2}{l}{\emph{Derived transit parameters}} \\
\hline
$a/R_{*}$ & $19.6 \pm 0.6$ &  $19.77_{-0.45}^{+0.38} $ \\
Stellar density $\rho_{*}$ [$ \rm g\;cm^{-3}$] & $1.79 \pm 0.17$ &  $1.85_{-0.12}^{+0.11} $ \\
Impact parameter $b$ & $0.14 \pm 0.14$ &  $0.13_{-0.08}^{+0.09} $ \\
Limb-darkening coefficient $u_{a}$  & $0.53 \pm 0.09$ &  $0.50 \pm 0.05 $ \\
Limb-darkening coefficient $u_{b}$  &  $0.13 \pm 0.26$  &  $0.09 \pm 0.12 $\\
& \\
\multicolumn{2}{l}{\emph{Atmospheric parameters of the star}} \\
\hline
Effective temperature $T_{\rm{eff}}$[K] & 5330 $\pm$ 120 & 5200 $\pm$ 100 \\
Spectroscopic surface gravity log\,$g$ [cgs]&  $4.55 \pm 0.14$ & 4.60  $\pm$ 0.15 \\
Derived surface gravity log\,$g$ [cgs]&  4.5  $\pm$ 0.1 &  4.50  $\pm$ 0.02 \\
Metallicity $[\rm{Fe/H}]$ [dex] & -0.07  $\pm$ 0.15 & 0.30  $\pm$ 0.12 \\
Stellar rotational velocity $V \sin{i_{*}}$ [\kms] & 3.5 $\pm$ 1.5 & 3.5 $\pm$ 1.5 \\
Spectral type & G8V & K0V \\
& \\
\multicolumn{2}{l}{\empty{Stellar and planetary physical parameters}} \\
\hline
Stellar mass [\Msun] &  $ 0.88 \pm 0.06 $ &  0.91 $\pm$ 0.04 \\
Stellar radius [\Rsun] &  $0.88 \pm 0.04$  &  $0.89 \pm 0.02$  \\
Planetary mass $M_{\rm p}$ [\Mjup ]  &  $9.9 \pm 0.5$ &  $10.1 \pm 0.4 $ \\
Planetary radius $R_{\rm p}$ [\Rjup]  &  $1.03 \pm 0.06$ &  $1.05 \pm 0.03 $ \\
Planetary density $\rho_{\rm p}$ [$\rm g\;cm^{-3}$] &  $11 \pm 2$ &  $11.0_{-0.9}^{+0.8}  $ \\
Planetary surface gravity log\,$g_{\rm p }$ [cgs] &  $ 4.36 \pm 0.03$ &  $4.36 \pm 0.03 $ \\
Age $t$ [Gyr]  & $6 \pm 3$ &  $6.2_{-2.8}^{+3.5} $ \\
Orbital semi-major axis $a$ [au] & 0.080 $\pm$ 0.005 &  0.0818 $\pm$ 0.0012 \\
Equilibrium temperature at the averaged distance $T_{\rm eq}$ [K] ~$^b$ & 850 $\pm$ 40 &  767 $\pm$ 16 \\
\hline       
\vspace{-0.5 cm}
\footnotetext[2]{Black-body equilibrium temperature assuming a null Bond albedo and uniform 
heat redistribution to the night side.}       
\end{tabular}
\end{minipage}
\label{table_param_KOI889}  
\end{table*}

\subsubsection{KOI-205}
As for Kepler-75, the SC orbital and physical parameters agree very well with those determined by \citet{Diazetal2013}
(see Table~\ref{table_param_KOI205}). 
The best fit of the SC transit and radial-velocity observations is shown in 
Fig.~\ref{fig_KOI205}.

\begin{figure*}[t]
\centering
\begin{minipage}{15cm}
%\vspace{-0.5cm}
%\includegraphics[width=7cm]{plot_transit_KOI205_SC_2014.pdf}
\includegraphics[width=7cm]{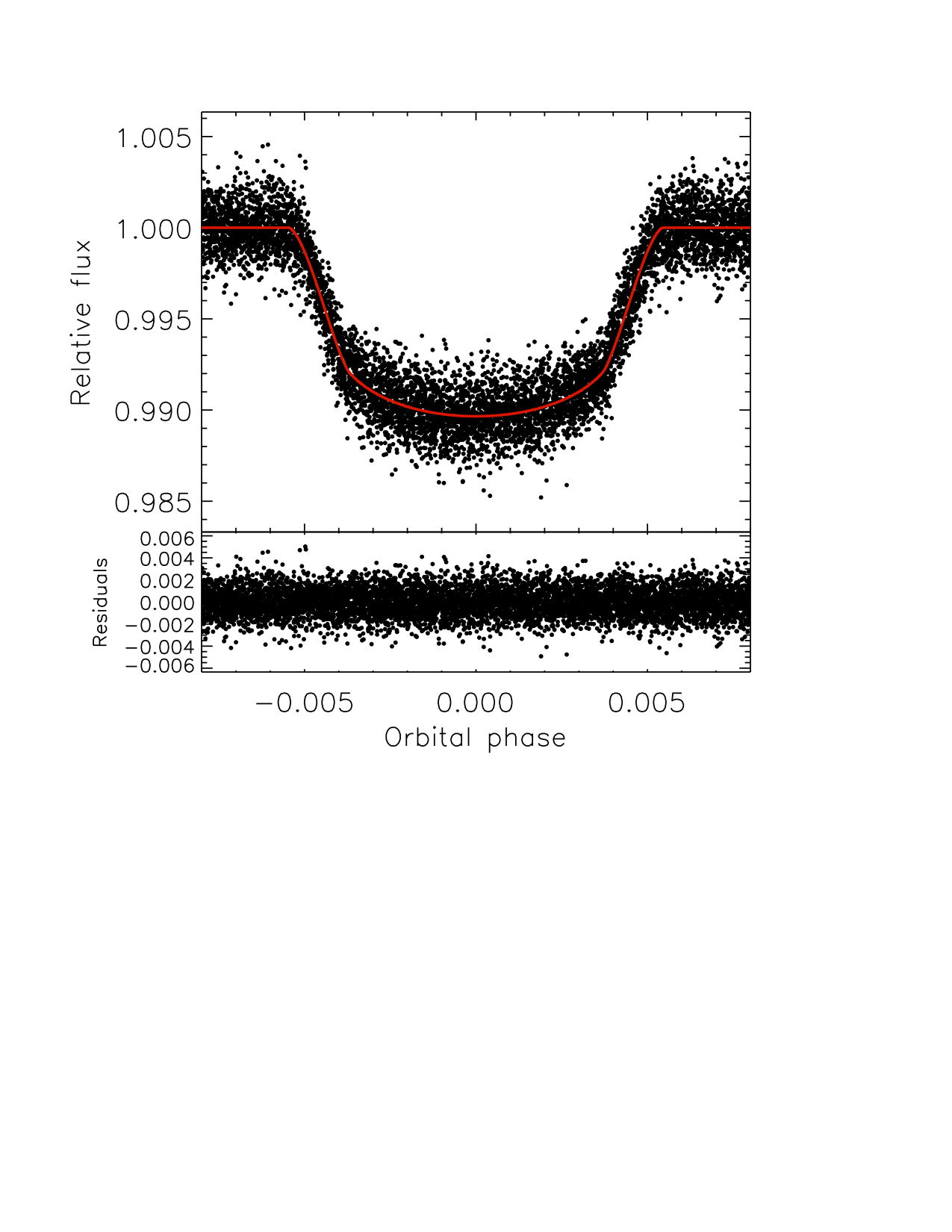}
\hspace{1.0cm}
\includegraphics[width=7cm]{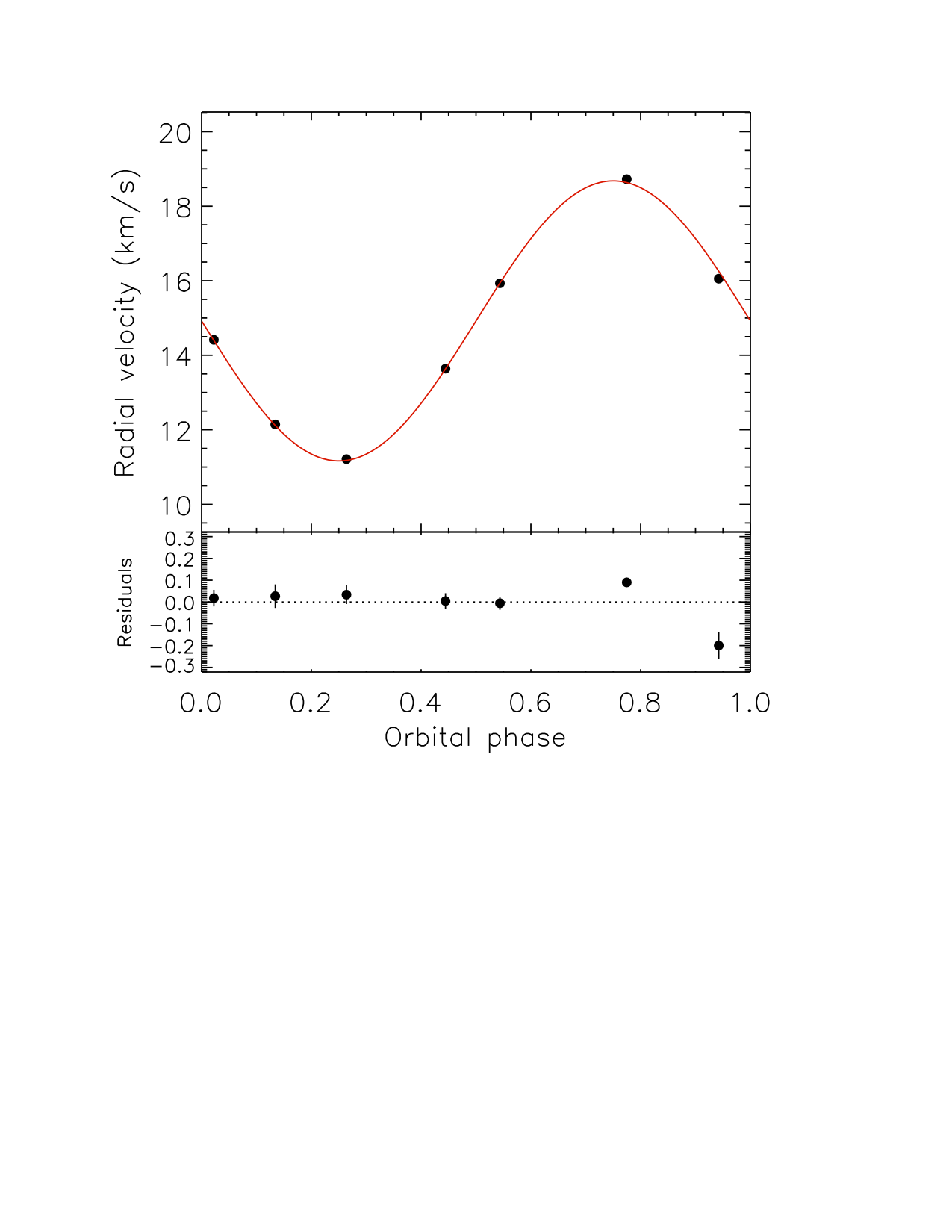}
\vspace{-3.7cm}
\caption{
\emph{Left panel}: phase-folded transit light curve of KOI-205 along with the transit
model (red solid line).
\emph{Right panel}: phase-folded radial-velocity curve of KOI-205 and, superimposed, the 
Keplerian model (red solid line).
}
\label{fig_KOI205}
\end{minipage}
\end{figure*}

\begin{table*}[h!]
\centering
\caption{KOI-205 system parameters.}            
\begin{minipage}[t]{13.0cm} 
\setlength{\tabcolsep}{1.0mm}
\renewcommand{\footnoterule}{}                          
\begin{tabular}{l l l}        
\hline\hline                 
\emph{Fitted system parameters}  & \citet{Diazetal2013} & This work\\
\hline
Orbital period $P$ [days] & 11.7201248 $\pm$ 0.0000021 &  $11.720126 \pm 0.000011 $\\
Transit epoch $T_{ \rm 0} [\rm BJD_{TDB}-2454900$] & 75.17325 $\pm$ 0.00012 &  $286.13609 \pm 0.00016 $ \\
Transit duration $T_{\rm 14}$ [h] & 3.07 $\pm$ 0.15 &  $3.069 \pm 0.017 $ \\
Radius ratio $R_{\rm p}/R_{*}$ & $0.09849 \pm 0.00049$ &  $0.09906 \pm 0.00094 $  \\
Inclination $i$ [deg] & $88.46 \pm 0.05$ &  $88.43 \pm 0.06 $ \\
Limb-darkening coefficient $q_{1}$  & - &  $0.45_{-0.09}^{+0.12} $ \\
Limb-darkening coefficient $q_{2}$  & - &  $0.34_{-0.15}^{+0.17} $  \\
$\sqrt{e}~\cos{\omega}$ & - &  $0.00 \pm 0.05 $ \\
$\sqrt{e}~\sin{\omega}$  & - &  $-0.03 \pm 0.010 $ \\
Orbital eccentricity $e$  &  $< 0.010$ &  $< 0.015 $ \\
Radial velocity semi-amplitude $K$ [\kms] & $3.732 \pm 0.039$ &  $3.757_{-0.084}^{+0.071} $ \\
Systemic velocity  $V_{\rm r}$ [\kms] & $15.057 \pm 0.026$ &  $14.922 \pm 0.050 $ \\
RV jitter [\ms] $s_{\rm j}$ & $\sim 40$ &  $120_{-50}^{+91} $ \\
& \\
\multicolumn{2}{l}{\emph{Derived transit parameters}} \\
\hline
$a/R_{*}$ & $25.20 \pm 0.42$ &  $25.07 \pm 0.43 $ \\
Stellar density $\rho_{*}$ [$ \rm g\;cm^{-3}$] & $2.18 \pm 0.10 $ & $2.17 \pm 0.11$\\
Impact parameter $b$ & $0.676 \pm 0.014$ &  $0.690 \pm 0.011 $ \\
Limb-darkening coefficient $u_{a}$  & $0.523 \pm 0.017$~$^a$ &  $0.47 \pm 0.17 $ \\
Limb-darkening coefficient $u_{b}$  & $0.187 \pm 0.011$~$^a$ &  $0.21 \pm 0.24 $ \\
& \\
\multicolumn{2}{l}{\emph{Atmospheric parameters of the star}} \\
\hline
Effective temperature $T_{\rm{eff}}$[K] & 5237 $\pm$ 60 & 5400 $\pm$ 75 \\
Spectroscopic surface gravity log\,$g$ [cgs]&  $4.65 \pm 0.07$   & 4.7 $\pm$ 0.1 \\
Derived surface gravity log\,$g$ [cgs]&  $4.550 \pm 0.015$   &  $4.558  \pm 0.014 $ \\
Metallicity $[\rm{Fe/H}]$ [dex] & 0.14  $\pm$ 0.12 & 0.18  $\pm$ 0.12 \\
Stellar rotational velocity $V \sin{i_{*}}$ [\kms] & 2.0 $\pm$ 1.0 & 2.0 $\pm$ 1.0 \\
Spectral type & K0V & G7V \\
& \\
\multicolumn{2}{l}{\emph{Stellar and planetary physical parameters}} \\
\hline
Stellar mass [\Msun] &  0.92 $\pm$ 0.03 &  $ 0.96_{-0.04}^{+0.03}  $ \\
Stellar radius [\Rsun]  &  $0.84 \pm 0.02$ &  $ 0.87 \pm 0.02 $  \\
BD mass $M_{\rm b}$ [\Mjup]  &  $39.9 \pm 1.0$ &  $40.8_{-1.5}^{+1.1} $ \\
BD radius $R_{\rm b}$ [\Rjup]  &  $0.81 \pm 0.02$ &  $0.82 \pm 0.02 $ \\
BD density $\rho_{\rm b}$ [$\rm g\;cm^{-3}$] &  $100.4 \pm 6.9$ &  $90.9_{-6.8}^{+7.2} $ \\
Age $t$ [Gyr]  & $3.1_{-1.3}^{+2.4}$ &  $1.7_{-1.2}^{+2.5} $ \\
Orbital semi-major axis $a$ [au] & 0.0987 $\pm$ 0.0013 &  $0.1010 \pm 0.0010 $ \\
\hline       
\vspace{-0.5cm}
\footnotetext[1]{The limb-darkening coefficients were allowed to vary within their $1\sigma$ errors 
related to the uncertainties on stellar atmospheric parameters.} 
\end{tabular}
\end{minipage}
\label{table_param_KOI205}  
\end{table*}

\section{Discussion and conclusions}
The analysis of \emph{Kepler} SC photometry, RV data, SOPHIE and 
ESPaDONs spectra of seven giant companions has permitted us
to refine their orbital and physical parameters.
In three cases, namely Kepler-43, Kepler-75, and KOI-205, 
they agree with published parameters within $1-1.3~\sigma$. 

For Kepler-44, for which only two quarters of LC 
data (Q1 and Q2) were analysed for the discovery announcement,
the new transit parameters determined with our DE-MCMC approach
indicate a transit duration shorter by $2~\sigma$, 
hence a slightly larger $a/R_\star$ and higher stellar density. 
This, in turn, implies that the host star
and its planetary companion are smaller than previously found. 

A separate discussion must be made for Kepler-39b and Kepler-74b 
because we have revised the significance of their 
orbital eccentricities. That of Kepler-39b is detected 
with a 2~$\sigma$ significance level and, according to the Lucy-Sweeney criterion,
it might be spurious. Slight asymmetries in the RV curve caused by 
residual effects from the correction of moonlight contamination \citep{Santerneetal2011b} 
and/or CCD charge transfer inefficiency \citep{Bouchyetal2009} 
might cause false eccentricities. Indeed, these
effects become strong when observing faint stars.
More radial-velocity observations without moonlight contamination
are required to determine whether Kepler-39b has a low eccentricity.
Kepler-74 would also benefit from additional high-accuracy and high-precision
RVs because the chains of our DE-MCMC combined analysis did not
converge towards a unique solution when including the eccentricity as a free parameter.
For this reason, we decided to fit only a circular model to \emph{Kepler} and RV data.
For circular orbits, both Kepler-39b and Kepler-74b 
would be smaller and denser than previously found. In any case,
the radius of Kepler-39b is still larger than predicted by theoretical isochrones (see Fig.~\ref{fig_MR_BD}),
as highlighted by \citet{Bouchyetal2011}. 
 
Our new spectral analyses of the available co-added spectra revealed
slightly hotter effective temperatures of Kepler-39 and KOI-205 and 
significantly higher metallicities for Kepler-39 and Kepler-75.

Among our seven targets, the most striking divergence with already published 
parameters was found for Kepler-41. Indeed, SC data point to a considerably 
lower inclination, higher impact parameter, and lower $a/R_\star$ 
than found by \citet{Santerneetal2011a} and \citet{Southworth2012}.
In consequence, both the host star and the planet have larger radii 
than previously derived. This new solution also has an impact on the 
estimation of the planetary geometric albedo that is 
significantly lower than previously estimated:  $A_{\rm g} < 0.135$ .
Both the larger radius and the lower albedo make this planet
resemble the majority of hot Jupiters. Conversely, the analysis of 
\emph{Kepler} LC data had erroneously resulted in  
peculiar characteristics for this planet. 
This emphasizes that, in some cases, 
SC data are necessary to derive accurate system parameters, 
in addition to reducing covariances between transit parameters \citep{PriceRogers2014} and
permitting to compute precise transit timing variations.
This will also be considered in the light of the future TESS and PLATO space missions.

\begin{acknowledgements}
A.~S.~Bonomo and R.~F.~D\'iaz acknowledge funding from the 
European Union Seventh Framework Programme (FP7/2007-2013) under Grant agreement number 313014 (ETAEARTH). 
A.~Santerne is supported by the European Union under a Marie Curie Intra-European Fellowship 
for Career Development with reference FP7-PEOPLE-2013-IEF, number 627202. 
This research has made use of the results produced by the PI2S2 Project managed by the Consorzio COMETA,
a co-funded project by the Italian Ministero dell'Istruzione, 
Universit\`a e Ricerca (MIUR) within the Piano Operativo Nazionale Ricerca Scientifica, 
Sviluppo Tecnologico, Alta Formazione (PON 2000Ð2006).
\end{acknowledgements}

~\\
~\\

\end{document}